\documentclass[twoside,twocolumn,aps,pra,longbibliography]{revtex4-1}
\usepackage[latin9]{inputenc}
\usepackage{graphicx}
\usepackage{amssymb}
\usepackage{amsmath}
\usepackage{amsfonts}
\usepackage{epsfig}
\usepackage{wrapfig}
\usepackage{refcount}
\usepackage{bm}
\usepackage{hyperref}
\usepackage{verbatim}
\usepackage[margin=1in,nofoot]{geometry}

\newcommand {\be}{\begin{equation}}
\newcommand {\ee}{\end{equation}}
\newcommand {\bea}{\begin{eqnarray}}
\newcommand {\eea}{\end{eqnarray}}

\newcommand{\cl}{\mathcal}
\usepackage[margin=1in,nofoot]{geometry}

\begin{document}

\title{Modular Quantum Information Processing by Dissipation}

\author{Jeffrey Marshall, Lorenzo Campos Venuti, Paolo Zanardi}

\affiliation{Department of Physics and Astronomy, and Center for Quantum Information
Science \& Technology, University of Southern California, Los Angeles,
CA 90089-0484}
\begin{abstract}
Dissipation can be used as a resource to control and simulate quantum systems. We discuss a modular model based on fast dissipation  capable of performing universal quantum computation, and simulating arbitrary Lindbladian dynamics. The model consists of a network of elementary dissipation-generated modules and it is in principle scalable.
In particular, we demonstrate the ability to dissipatively prepare all single qubit gates, and the CNOT gate; prerequisites for universal quantum computing. We also show a way to implement a type of quantum memory in a dissipative environment, whereby we can arbitrarily control the loss in both coherence, and concurrence, over the evolution. Moreover, our dissipation-assisted modular construction exhibits a degree of inbuilt robustness to Hamiltonian, and indeed Lindbladian errors, and as such is of potential practical relevance.

\end{abstract}
\maketitle
\section{Introduction}
 A long-standing  view held in the Quantum Information Processing (QIP) community is that all noise is detrimental to the performance of a quantum system in which QIP is carried out. This view-point prompted a large research effort into the development of active quantum error correction (QEC) \cite{Lidar-Brun:book,Gottesman:1996fk,Kielpinski:01} as well as  into passively protecting quantum states from noise. Passive schemes, include the use of decoherence-free subspaces \cite{Zanardi:97c,Lidar:1998fk} and noiseless subsystems \cite{Knill:2000dq,Zanardi:99d,Zanardi:2003c}.

More recently the QIP community is becoming increasingly aware that noise, either environmental or artificial, can be harnessed and exploited  to perform QIP primitives. For example  dissipation can be turned into a {\em{resource}} for the preparation of entangled states \cite{Kraus-prep}, to enact quantum simulation \cite{barreiro2011open}, and quantum annealing schemes \cite{childs:01,sarandy:05}, to name just a few.

In this paper we build upon the techniques developed in \cite{zanardi-dissipation-2014,dissi_2nd_order:Zanardi2016}, whereby information can be encoded in the steady states of a strongly dissipative system. There are in-fact two regimes for this so-called dissipation theorem; 1) The full dissipative system is equivalent (up to boundable errors) to a coherent, unitary one (Sect. \ref{sect:coherent} below) and 2) where the system can simulate incoherent (i.e., Lindblad) dynamics (Sect. \ref{sect:incoherent} below). We will discuss in the next section the assumptions which govern whether case 1 or case 2 is achieved.

The power of this method is derived from the fact that the dissipation drives the system naturally to the steady state manifold while at the same time renormalizing Hamiltonian terms. In fact, it was shown in \cite{zanardi-dissipation-2014}, that at first order in the dissipation theorem (i.e., case 1 above), that the application of a `small' control Hamiltonian (on top of the dominant dissipative dynamics) can be used to coherently traverse a path through such a set of steady states (SSS). Though errors do accumulate, these are shown to be proportional to the dissipative time-scale and can be made arbitrarily small in the fast dissipation regime. In short, up to boundable errors, and some appropriate assumptions, the full dissipative system can be shown to be equivalent to a unitary dynamics in the SSS.

In the following, we shall give a theoretical outline for a scalable architecture which can be used to enact universal quantum computation, and also simulate arbitrary Lindblad type systems \cite{dissi_2nd_order:Zanardi2016}. The system comprises  a network of so called dissipative-generated modules (DGMs). Couplings between the DGMs define the global dynamics and can be tailored  to enact universal QIP.  Interestingly, certain types of errors, both at the Hamiltonian control level and in the noise model, are suppressed. This endows our strategy with some degree of inbuilt robustness.

We will first overview the relevant theoretical background in Sect.~\ref{sect:background}, followed by outlining the specific construction of a DGM network in Sect.~\ref{sect:network}. After this, we provide several insightful examples of these networks, in both the coherent (Hamiltonian), and incoherent (Lindbladian) cases, in Sect.~\ref{sect:examples}. Perhaps our main relevant example is the ability to generate two qubit entangling gates, and indeed a CNOT gate, stemming from an effective Hamiltonian generator, Eq.~\eqref{eq:entangling}. Before concluding, we show that a DGM network is indeed highly resilient to certain types of encoding error, both Hamiltonian and dissipative, and we provide numerical examples of this robustness, in Sect.~\ref{sect:robustness}.

\section{Preliminaries}\label{sect:preliminaries}
\subsection{Background}\label{sect:background}
Here we briefly summarize the main results of Refs.~\cite{zanardi-dissipation-2014} and \cite{dissi_2nd_order:Zanardi2016}. We denote by $\cl{H}$, the Hilbert space of the system, and we assume a time-independent Liouvillian super-operator, $\cl{L}_0$, acting on the algebra of linear operators over $\cl{H}$ (denoted L$(\cl{H})$). We define the set of steady states (SSS) as the set of quantum states $\rho \in \text{L}(\cl{H})$ (i.e.~$\rho \ge 0,\,\text{Tr}(\rho)=1$) such that $\cl{L}_0 (\rho)=0$.  We call  $\cl{P}_0$ the spectral projection of $\cl{L}_0$ associated to eigenvalue zero. Under the assumptions that $e^{t\cl{L}_0}$ is a completely positive trace-preserving (CPTP) map ($t\ge 0)$, because of von Neumann mean ergodic theorem for contraction semigroups (see e.g.~\cite{sato_ergodic_1977}), we have $\cl{P}_0=\overline{e^{t \cl{L}_0}}$ (with $\overline{f} = \lim_{T\to \infty} T^{-1} \int_0^T f(t) dt$), implying that also $\cl{P}_0$ is CPTP, being a convex combination of CPTP maps. Moreover one can show that $\cl{P}_0\cl{L}_0=\cl{L}_0\cl{P}_0=0$ i.e., there is no nilpotent term associated to the zero eigenvalue (see e.g.~\cite{venuti_adiabaticity_2016}). We define the reduced resolvent of $\cl{L}_0$ as $\cl{S}=\lim_{z\rightarrow 0} \cl{Q}_0 (\cl{L}_0-z)^{-1}\cl{Q}_0$, where $\cl{Q}_0:=1-\cl{P}_0$ is the projector onto the complementary subspace of ker($\cl{L}_0$). The dissipative time-scale $\tau$ associated with this dynamics is given by  $\tau^{-1}:=\min_{\lambda \ne 0} |\Re \lambda|$ ($\lambda$ eigenvalues of $\cl{L}_0$), i.e.~$\| \cl{S} \| = O(\tau)$.
  We may add to $\cl{L}_0$ a Hamiltonian term, $\cl{K}=-i[K,\cdot]$ (we set $\hbar =1 $ for convenience), where $K=K^\dagger$, such that the full dynamics is governed by $\cl{L}=\cl{L}_0+\cl{K}$, i.e.,
 \be
 \frac{d \rho (t)}{dt} = \cl{L}(\rho(t)).
 \ee
\par
In Ref.~\cite{zanardi-dissipation-2014} it was shown that the evolution  under $\cl{L}$ inside the SSS can be approximated by an effective Hamiltonian generator of the form 
\be
\label{eq:effectiveHam}
\cl{K}_{eff}:= \cl{P}_0\cl{K}\cl{P}_0.
\ee
More specifically one has:
\be
\label{eq:dissi}
\|(\cl{E}_T-e^{\cl{\tilde{K}}_{eff}})\cl{P}_0\|=O(\tau/T),
\ee
where the exact dynamics are given by $\cl{E}_T:= e^{T \cl{L}}$, and $\cl{\tilde{K}}_{eff} := \cl{P}_0\cl{\tilde{K}}\cl{P}_0$, with $\cl{K}=T^{-1}\cl{\tilde{K}}$ and $\|\cl{\tilde{K}}\|=O(1)$.
Indeed, using the above result it was shown in \cite{zanardi-dissipation-2014}, for example,  that one can perform any single qubit operation in a strongly dissipative four qubit system. 
\par
The generator in Eq.~\eqref{eq:effectiveHam} will be referred to as the dissipation projected generator. 
If it happens that $\cl{P}_0\cl{K}\cl{P}_0=0$, the effective evolution is then controlled by
\be
\label{eq:2ndOrderGenerator}
\cl{L}_{eff}:=-\cl{P}_0\cl{KSKP}_0.
\ee
In this case one obtains 
\be  
\label{eq:2ndOrder}
\| (\cl{E}_T-e^{\cl{\tilde{L}}_{eff}})\cl{P}_0\|=O(\sqrt{\tau/T})
\ee
where $\cl{\tilde{L}}_{eff}=-\cl{P}_0\cl{\tilde{K}S\tilde{K}P}_0$ and now $\cl{K}=\cl{\tilde{K}}/\sqrt{T}$ \cite{dissi_2nd_order:Zanardi2016}. 
In this context dissipation becomes a \emph{resource} which allows one to simulate arbitrary Lindblad dynamics \cite{dissi_2nd_order:Zanardi2016}.

\subsection{DGM Network}\label{sect:network}
We propose a new computation/simulation paradigm using the dissipation projected dynamics as overviewed in the previous subsection. In particular, we construct a network of so called dissipation-generated modules (DGMs), where the full dynamics within the network is determined by couplings between these dissipative modules.
\par
\begin{figure}[h]
\vspace{-4mm}
\begin{centering}
\includegraphics[scale=0.3]{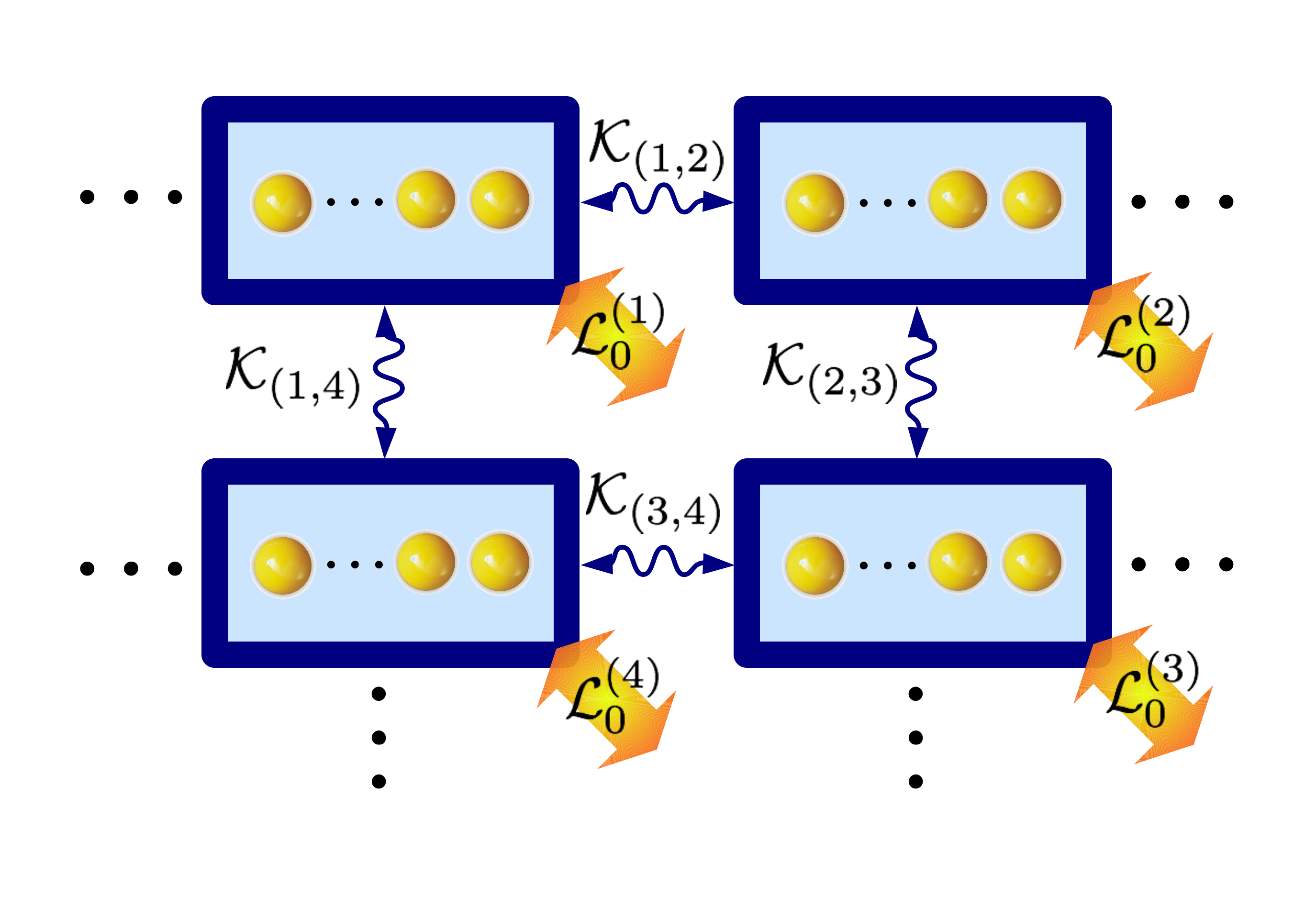}
\end{centering}
\vspace{-8mm}
\caption{(Color online) A four-module patch of a two-dimensional network of coupled DGMs. The $i$-th module dissipates according the Liouvillian ${\cal L}_0^{(i)}$
and the bare inter-module coupling are given by the ${\cal K}_{(i,j)}=-i[K_{(i,j)},\cdot]$. Yellow (gray) thick  arrows represent strong dissipation whereas blue (black)
wavy lines represent Hamiltonian interactions.}
\label{DGM-fig}
\end{figure}

The Hilbert space of a single dissipation-generated module is of the form $\cl{H}=\cl{H}_{1/2}^{\otimes n} \otimes \cl{H}_a$, where $\cl{H}_{1/2}$ is the Hilbert space of a single qubit, and $\cl{H}_a$ the Hilbert space of dissipative ancillary resources (e.g. qubits, bosonic modes, etc.). By allowing a coupling between different modules, one can construct a `DGM-network', which defines an undirected graph $\cl{G}:=(V,E)$, where the vertices $V$ are distinct modules, and the edges $E$, are defined by the couplings between modules. The full Hilbert space is therefore $\cl{H}=\otimes_{i\in V}\cl{H}_i$, where $\cl{H}_i$ has the same structure as above. This structure implies that the full Hamiltonian can be written in general as 
\be
\label{DGM_ham}
K=\sum_{(i,j)\in E} K_{(i,j)}.
\ee
Moreover, each module is assumed to be dissipative, such that the unperturbed dynamics of the system may be described by a Liouvillian super-operator of the form $\cl{L}_0=\sum_{i\in V} \cl{L}_0^{(i)}$, where $\cl{L}_0^{(i)}$ only acts on the $i$th module. With each $\cl{L}_0^{(i)}$ we associate a dissipative time-scale, $\tau_i$. The overall, dissipative, timescale of the whole network is given by $\tau:=\max_{i}( \tau_i$). We illustrate a four module patch of a two-dimensional DGM-network in Fig.~\ref{DGM-fig}.

Since the dissipation acts, by definition, separately on each module, it is clear that the full SSS projection operator over an entire DGM network is simply of the form $\cl{P}_0=\otimes _{i\in V}\cl{P}_0^{(i)}$, where $\cl{P}_0^{(i)}$ is the projection operator associated with the $i$th module. The Hamiltonian super-operator, $\cl{K}=\sum_{(i,j)\in E}\cl{K}_{(i,j)}$, gives rise to first-order effective dynamics as defined by Eq.~\eqref{eq:effectiveHam}, given by
\be
\label{eq:keff}
\cl{K}_{eff}=\sum_{(i,j)\in E} (\cl{P}_0 ^{(i)}\otimes \cl{P}_0^{(j)})\cl{K}_{(i,j)} (\cl{P}_0 ^{(i)}\otimes \cl{P}_0^{(j)}).
\ee
Errors associated with this effective dynamics, as per Eq.~\eqref{eq:dissi}, are of the order $\epsilon=O(J_{MAX} |E|\tau)$, where $J_{MAX}$ is the maximum inter-module coupling strength, and $|E|$ the number of edges in the network \footnote{This estimate is a worst case scenario which assumes that all the couplings $K_{(i,j)} $are always on}.

This error presents a trade-off between the graph complexity (as measured by $|E|$), and the global dissipation rate, $\tau^{-1}$. This trade-off stems from the observation that simple networks of $N$ nodes, e.g. a tree with $|E|=O(N)$, may require a greater number of elementary gates to perform some computation as compared to a fully connected graph with $|E|=O(N^2)$, indicating there will be an optimal graph configuration $\cl{G}$ for a given simulation. 
For arbitrary computations/simulations, one needs to be able to connect any two nodes in $\cl{G}$, where the cost of doing so is equal to the distance between these nodes (i.e.~this is the order of the number of gates one would need to apply). 

Finally note that the dissipation required for these modules can in principle be engineered using the second-order dissipation lemma  as given by 
Eq.~\eqref{eq:2ndOrder} (see  \cite{dissi_2nd_order:Zanardi2016}). 


\section{Examples}\label{sect:examples}
\subsection{Coherent Dynamics}\label{sect:coherent}
If the type of noise acting on the system admits a Decoherence-Free Subspace (DFS) in each module, Eq.~\eqref{eq:keff} becomes
\be
\label{eq:dfs}
K_{eff}=\sum_{i,j}(\Pi_i\otimes \Pi_j) K_{(i,j)} (\Pi_i \otimes \Pi_j),
\ee
where $\Pi_i$ is the orthogonal projector over the DFS in module $i$ \cite{zanardi-dissipation-2014,Albert:geometry_response}. 
\par
As an example, assume two connected DGMs, of $N$ qubits, each module undergoing collective amplitude damping of the form 
\be
\label{eq:diss}
\cl{L}_0^{(i)}(\rho)=\tau^{-1}_i( S^-\rho S^+ -\frac{1}{2}\{S^+ S^-,\rho \})
\ee
 where $i=1,2$ indicates the DGM index. The Hilbert space for $N$ qubits breaks down into separate angular momentum sectors as $\cl{H}=\oplus_J n_J \cl{H}_J$, where $n_J$ is the multiplicity of $\cl H_J=\text{span}\{|J,m\rangle\}_{m=-J}^J$. Under this type of dissipation, there is a DFS containing the lowest-weight angular momentum vectors in each sector (i.e.~$|J,-J\rangle$). Let us concentrate on the case $N=2$ in the following. In this case  there is a two-dimensional DFS (i.e~a qubit), spanned by the singlet, $|\bar 0 \rangle:=|0,0\rangle$, and the $|\bar 1\rangle := |1,-1\rangle$ triplet state \footnote{\unexpanded{$|0,0\rangle=\frac{1}{\sqrt{2}}(|01\rangle-|10\rangle),|1,-1\rangle=|00\rangle$}.}. The projectors introduced above are then given by $\Pi_i =|\bar 0 \rangle \langle \bar 0| + |\bar 1\rangle \langle \bar 1|$. This set-up in fact allows us to generate effective, logical, single qubit gates, as well as entangling gates.
 \par
Consider the two-local Hamiltonian $K=g\sqrt{2} \, \mathbb{I} \otimes \sigma^x$. 
From Eq.~\eqref{eq:dfs} one obtains
 \be
 K_{eff}=g \bar \sigma ^x := g (|\bar 1 \rangle \langle \bar 0 |+|\bar 0 \rangle \langle \bar 1| ),
 \ee
 i.e., a logical, effective, Pauli $x$ Hamiltonian.
 \par
 In a similar manner, $K=g \sigma^z \otimes \sigma ^z$ can be used to generate a logical Pauli $z$,  $\bar \sigma^z:=|\bar 1 \rangle \langle \bar 1 |-|\bar 0 \rangle \langle \bar 0|$. $\bar \sigma^y$, of course, can be generated using the commutation relations \footnote{Consider the Hamiltonian \unexpanded{$K=\sigma^z \otimes \sigma ^z$}. Clearly the action of this on \unexpanded{$|\bar 1\rangle$} is the identity, whereas results in a minus sign on \unexpanded{$|\bar 0\rangle$}. Therefore, this will result in a logical \unexpanded{$\sigma^z$} Hamiltonian. \unexpanded{$\bar \sigma^x$} is derived in a similar manner.}. Hence using this type of dissipation, one can generate all of the single qubit gates required for universal quantum computation.
 \par
\begin{figure}[h]
\begin{centering}
\includegraphics[scale=0.4]{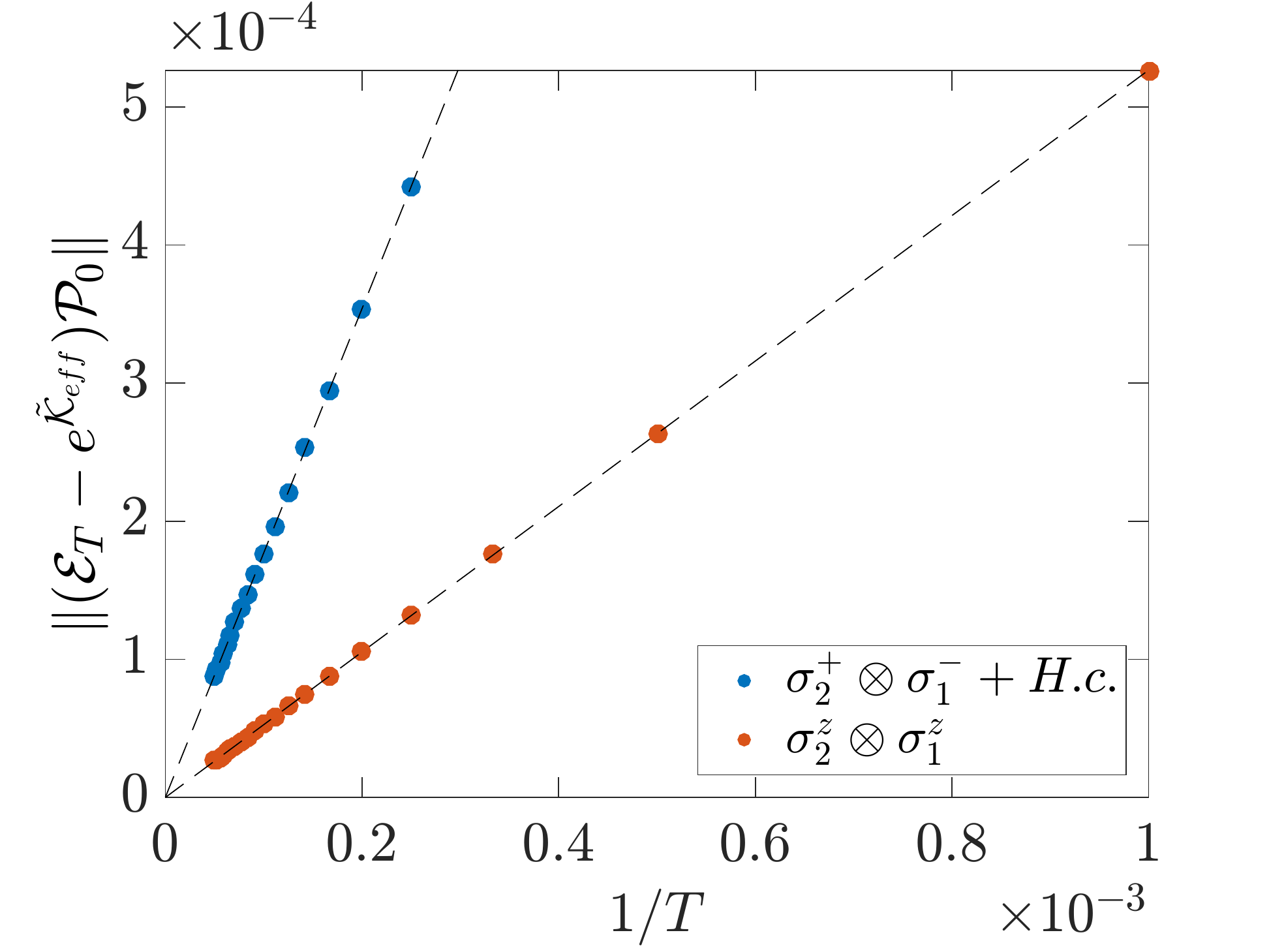}
\end{centering}
\caption{(Color online) Two-module network. Distance between the exact evolution $\cl{E}_T$ after time $T$, and the effective one for the two Hamiltonians discussed in the main text, $K=g( \sigma_2^+ \otimes \sigma_1^- + \sigma_2^- \otimes \sigma_1^+)$ [blue/dark gray circles], and $K=g\, \sigma_2^z \otimes \sigma_1^z$ [red/light gray circles]. We picked the dissipation time-scales to be: $\tau_1=1$ arb.~units, $ \tau_2=0.5$ arb.~units. We set $gT=1$. The linear fit is obtained using the least-squares fitting on all of the data points, and the norm is the maximum singular value of the maps realized as matrices. Time is measured in arbitrary units.}
\label{coherent-fig}
\end{figure}

 We now study the coupling between the two DGMs defined above. In particular consider the following interaction 
 \be
 \label{2dgm_ham}
 K=g(\sigma_2^+ \otimes \sigma_1^- + \sigma_2^- \otimes \sigma_1^+),
 \ee
where the tensor product is the product between the DGMs and the sub-index labels the qubit inside each DGM. 
According to Eq.~\eqref{eq:dfs} the above term induces the following effective Hamiltonian
\be
\label{eq:entangling}
K_{eff}=\frac{g}{2}(|\bar 0 \bar1 \rangle \langle \bar 1 \bar 0 | + |\bar 1 \bar 0 \rangle \langle \bar 0 \bar 1|). 
\ee
The interaction in Eq.~\eqref{eq:entangling} can be used to generate entanglement between DGMs. In fact, this can be used to create a `square root of $i$ swap' gate, SQ$i$SW, which in turn can be used to generate a CNOT gate \cite{bialczak:q_tomography} (see Appendix~\ref{app:cnot} for an explicit construction). A coupling of $K=g\, \sigma_2^z \otimes \sigma_1^z$ induces an effective generator of the form $K_{eff}=g|\bar 0 \bar 0 \rangle \langle \bar 0 \bar 0|$, which can also be used to generate entanglement between two modules. We provide a brief numerical verification of these two effective dynamics in Fig.~\ref{coherent-fig}.

Given that we can, in a similar fashion, generate all of the  single qubit gates, this method can be used to enact universal QIP over the logical of the modules. In Sect.~\ref{sect:robustness} we will show how this system is robust to certain types of Hamiltonian, and Lindbladian, encoding errors, therefore suggesting that DGM networks with couplings of type Eq.~\eqref{2dgm_ham} are an attractive arena to perform quantum information processing tasks.
\par

We finally consider an example with the same collective dissipation as in Eq.~\eqref{eq:diss} with $N=3$, i.e.~each DGM contains three qubits.  
In this case the eight dimensional Hilbert space of each module  $\cl{H}_{1/2}^{\otimes 3}=2\cl{H}_{1/2} \oplus \cl{H}_{3/2}$ has a three-dimensional DFS ($=\text{span}\{|\bar 0\rangle, |\bar 1\rangle, |\bar 2\rangle\}$) \footnote{We define \unexpanded{$|\bar 0\protect\rangle := |1/2,0,-1/2\protect\rangle,\, |\bar1\protect\rangle:=|1/2,1,-1/2\protect\rangle,\, |\bar 2 \protect\rangle := |3/2,-3/2 \protect\rangle$}, where the second index on the first two states indicates which of the two spin 1/2 sectors it belongs. Qubit notation: \unexpanded{$|1/2,0,-1/2\protect\rangle = \frac{1}{\sqrt{2}}(|010\protect\rangle - |100\protect\rangle ),\, |1/2,1,-1/2\protect\rangle =\frac{1}{\sqrt{6}}(|100\protect\rangle + |100\protect\rangle-2|001\protect\rangle ),\, |3/2,-3/2\protect\rangle=|000\protect\rangle$}.}. We consider a similar coupling as before, $K=g(\sigma_3^+ \otimes \sigma_1 ^- + \sigma_3^-\otimes \sigma_1^+)$, i.e., a hopping term between two nearby qubits in the different modules. Such a term induces the following effective Hamiltonian
\be
K_{eff}=\frac{g}{\sqrt{3}}(|\bar 2 \bar 0 \rangle \langle \bar 1 \bar 2 | + |\bar 1 \bar 2 \rangle \langle \bar 2 \bar 0 | )-\frac{g}{3}(|\bar 2 \bar 1 \rangle \langle \bar 1 \bar 2 |+|\bar 1 \bar 2 \rangle \langle \bar 2 \bar 1 |),
\ee
namely a coupling between all three logical levels is induced.

\subsection{Incoherent DGMs}\label{sect:incoherent}
From Eq.~\eqref{eq:2ndOrder}, if it is the case that $\cl{P}_0\cl{KP}_0=0$, the resulting dynamics are second order, and in general dissipative,  with the effective generator given by Eq.~\eqref{eq:2ndOrderGenerator}. In particular it is possible to simulate any Lindblad type dissipator over a DGM of the form (see Sec.~III of Ref.~\cite{dissi_2nd_order:Zanardi2016})
\be
\label{eq:lindblad}
\cl{L}(\rho) = \sum_{k}\tau^{-1}_k( L_k \rho L_k^\dagger -\frac{1}{2}\{L_k ^\dagger L_k , \rho\}).
\ee
Adapting those results to a network of modules, this means we can easily simulate any Lindblad system of the form $\cl{L}=\sum_i \cl{L}^{(i)}$, where the sum is over modules. In the following we build upon the examples already discussed in Ref.~\cite{dissi_2nd_order:Zanardi2016}, but consider the new ingredient provided by the interaction between different modules. 

\subsubsection{Interacting Jaynes-Cummings cavities}
In Refs.~\cite{sete_quantum_2015,man:2015_cavity} it was observed that by coupling a qubit in a (lossy) cavity interacting with a bosonic mode, to an empty cavity supporting another bosonic mode, one can extend the coherence time of the qubit. By further adding an atom (qubit) in the extra cavity, it was also shown that the entanglement between the atoms survives for longer times. Similar results have also been found in \cite{gonzalez_coherence}.

We follow a more general set-up, of which the system described in \cite{man:2015_cavity} is a special case. 
We are able to show that by increasing the coupling strength between the modes, we can increase the effective dissipative time-scale of the qubits, hence increasing the time window where purely quantum effects can be observed. 

\begin{figure}[h]

\begin{centering}
\includegraphics[scale=0.3]{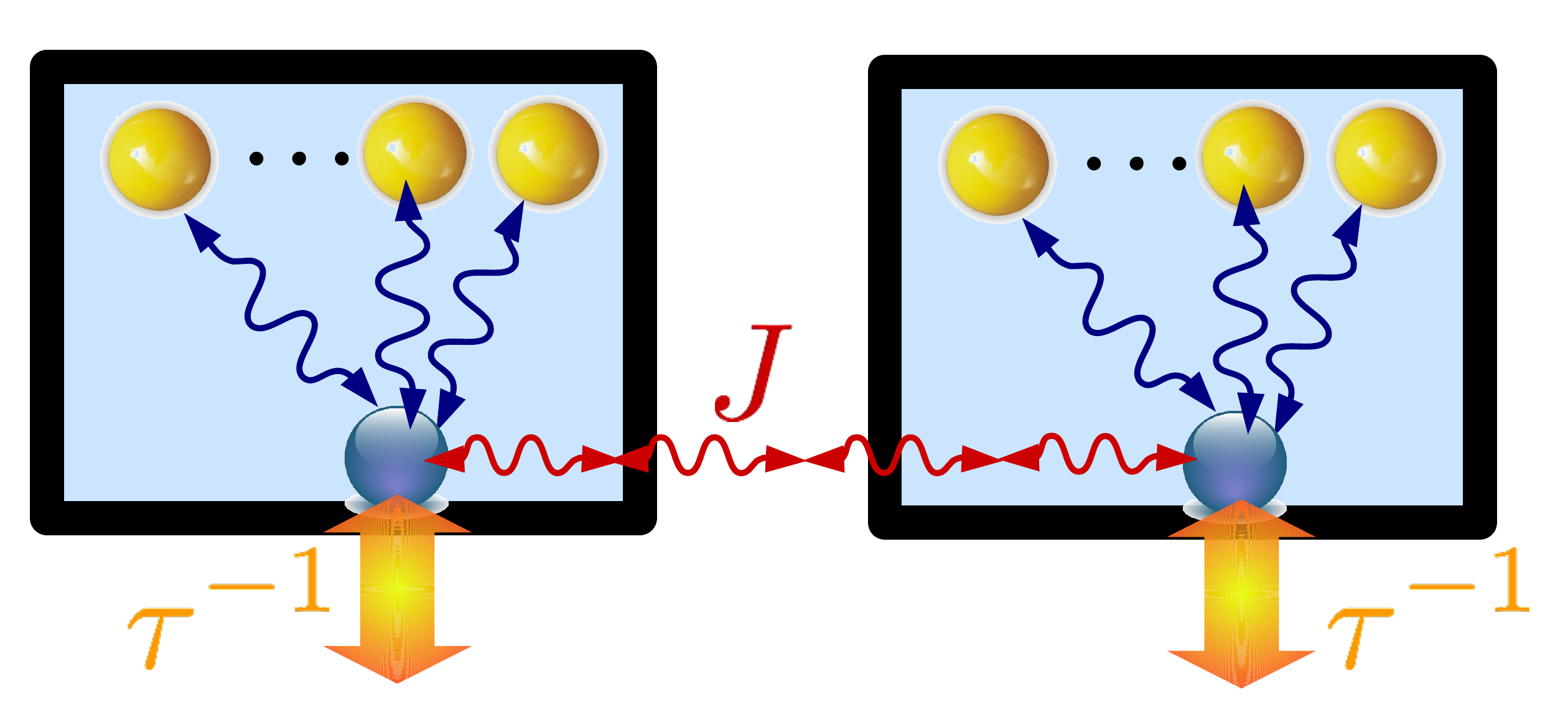}
\end{centering}

\caption{(Color online) Two modules, each containing qubits (yellow/light gray spheres, top) connected to a  bosonic mode (blue/dark gray sphere, bottom), which dissipate at rate $\tau^{-1}$. In turn, the two modules are coherently coupled via the bosonic modes, with strength $J$.}
\label{DGM-fig2}
\end{figure}
\par
We consider $N_\alpha$ qubits ($\alpha=A,B$) in each cavity, collectively coupled to a single bosonic mode in each cavity via a Jaynes-Cummings term. 
The unperturbed generator for two modules is given by (see Fig.~\ref{DGM-fig2} for reference)

\be
\label{eqn_AB}
\cl L_0=-i[H_{AB},\cdot]+\sum_{\alpha=A,B}\cl L_0^{(\alpha)},
\ee 
where 
\be
\label{diss_AB}
\cl L_0^{(\alpha)}(\rho)= \tau^{-1}(c_\alpha \rho c^\dagger _\alpha -\frac{1}{2}\{c^\dagger_\alpha c_\alpha,\rho\}),
\ee
 and $c_\alpha (c_\alpha^\dagger)$ is the annihilation (creation) operator for mode in cavity $\alpha=A,B$.
 The coupling Hamiltonian is given by $H_{AB}=J(c_A^\dagger  c_B + c_A  c_B^\dagger)$. The unique steady state in the bosonic sector is the joint vacuum state $\rho=|0\rangle \langle 0| \otimes |0 \rangle \langle 0|$. The decay of the qubits is assumed to be mediated by the usual  Jaynes-Cummings interaction between the qubits and modes, i.e., 
\be
\label{qq_ham}
K=K_0+ \sum_{\alpha=A,B}g_\alpha(c_\alpha S_\alpha^+ + c_\alpha^\dagger S_\alpha^-),
\ee 
where $K_0=\sum_{\alpha=A,B}(\omega^q_\alpha S_\alpha^z+ \omega_\alpha c_\alpha^\dagger c_\alpha)$. One can check the second term in Eq.~\eqref{qq_ham} vanishes at first order. For the sake of  simplicity we  set $\omega_\alpha =\omega^q_\alpha=0$ (see Appendix~\ref{sect:derivation}). 

This system is equivalent, up to arbitrary controllable errors to two coupled modules  containing $N_\alpha$  qubits per module (bosons frozen at $|0\rangle$). The effective dynamics are governed by Liouvillian
\be
\label{qq_eff}
\cl L_{eff} = -i[K_{eff},\cdot]+\sum_{\alpha=A,B}\cl L_{eff}^{\alpha},
\ee
with 
\be
K_{eff}=J_{eff}(S_A^+ S_B^- + S_A^- S_B^+),
\ee 
and
\be
\label{eff-diss}
\cl L_{eff}^{\alpha}(\rho_\alpha) = \tau^{-1}_{eff,\alpha}(S_\alpha^- \rho S_\alpha^+ -\frac{1}{2}\{S_\alpha^+ S_\alpha^-,\rho\}),
\ee
 where $\rho_\alpha$ is the $N_\alpha$-qubit state. The effective dissipative rate is found to be (see Appendix~\ref{sect:derivation} for derivation),
\be
\tau^{-1}_{eff,\alpha}:=\frac{4\tau}{1+4(J\tau)^2}g_{\alpha}^2,
\ee
 and the coupling strength,
 \be
 J_{eff}:=\frac{-4Jg_Ag_B\tau^2}{1+4(J\tau)^2}.
 \ee
\par
We observe immediately that the effective dissipative timescale is controlled by a factor of $J^2 \tau$. Also note that if we set $J=0$, we recover the result in Eq.~(5) of \cite{dissi_2nd_order:Zanardi2016}. 
\par
\begin{figure}[h]
\includegraphics[scale=0.4]{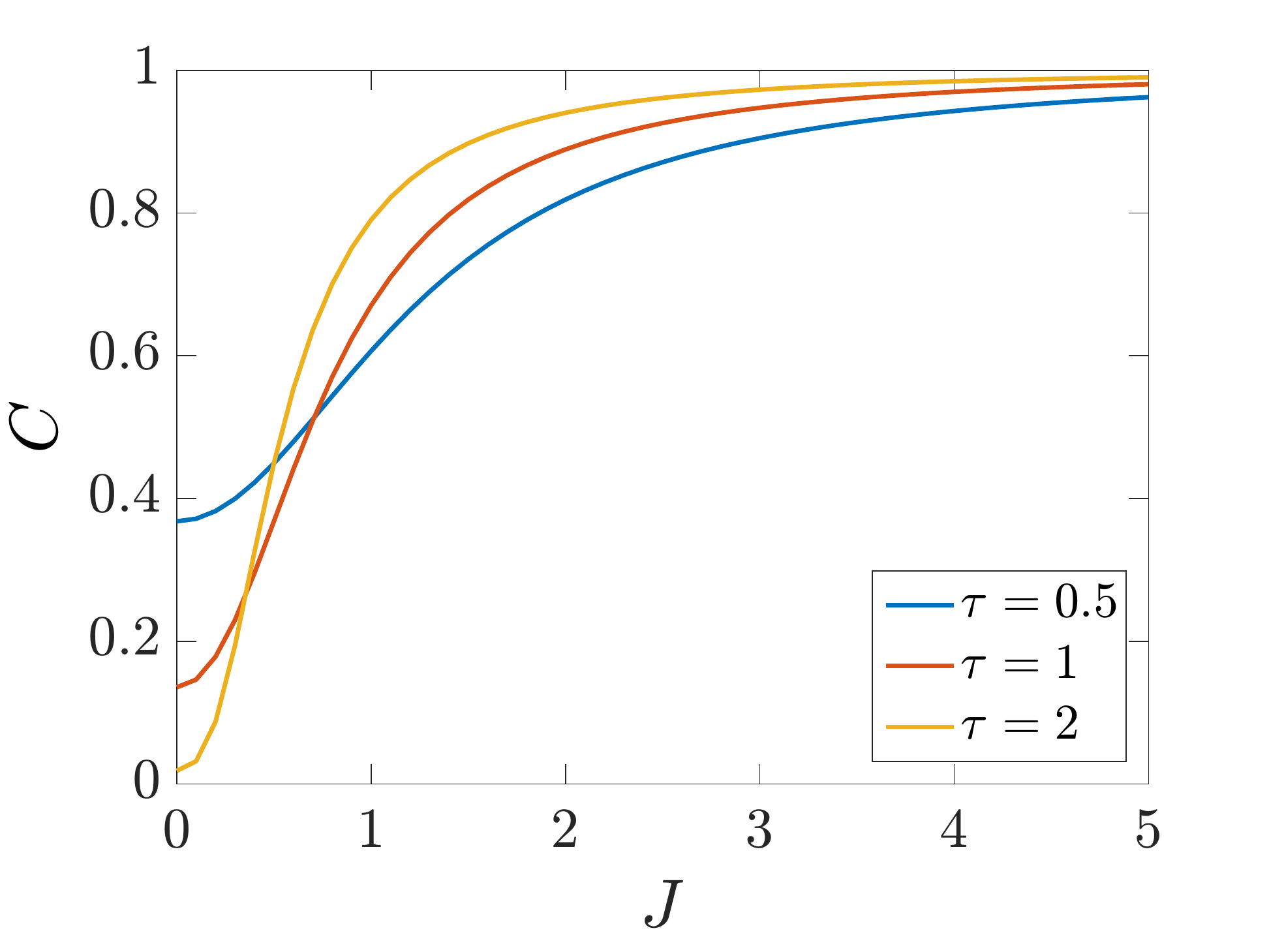}
\caption{(Color online) Coherence of the evolved effective dynamics, Eq.~\eqref{qq_eff}, as a function of $J$ for a single qubit, for three values of $\tau$ (see legend). Initial qubit state $|\psi_0\rangle =\frac{1}{\sqrt{2}}(|0\rangle +|1\rangle)$. We set $Tg_A^2=1$ arb.~units.  This result is valid only in the regime $T \gg \tau$. Time is measured in arbitrary units ($J,g_\alpha$ are inverse time).}
\label{fig:coherence}
\end{figure}

If we allow the second DGM to contain no qubit (e.g.~$g_B=0$), and place a single qubit in system $A$, we have the exact case as discussed in \cite{man:2015_cavity} (see Fig. 1 of \cite{man:2015_cavity}). The amount of coherence in a state $\rho$, in a given basis $i,j$, can be captured by $C=\sum_{i\neq j} |\rho_{ij}|$ \cite{man:2015_cavity}. 
It can be shown that (see \footnote{One can calculate the functional form of the coherence, in the computational basis (\unexpanded{$|i\rangle$}, $i=0,1$), by noting that \unexpanded{$\cl{L}_{eff}^\alpha (|i\rangle _\alpha \langle j |) = -\frac{\tau_{eff,\alpha}^{-1}}{2}|i\rangle _\alpha \langle j|$}, for $i\neq j$, and \unexpanded{$\cl{L}_{eff}^\alpha (|1\rangle _\alpha \langle 1|) = \tau_{eff,\alpha}^{-1}(|0\rangle _\alpha 0| - |1 \rangle _\alpha \langle 1|)$. Note, $|0\rangle _\alpha \langle 0|$} is a steady state. The subscript $\alpha$ on the states just indicate the system $\alpha=A,B$.}), in the standard basis, using the effective dynamics, one obtains $C=e^{-T\tau_{eff,A}^{-1}/2}$ [note that this result is correct up to $O(\sqrt{\tau/T})$]. 
A plot of $C$ as a function of $J$ for different dissipations, is given in Fig.~\ref{fig:coherence}. 

\par
\begin{figure}[h]
\begin{centering}
\includegraphics[scale=0.4]{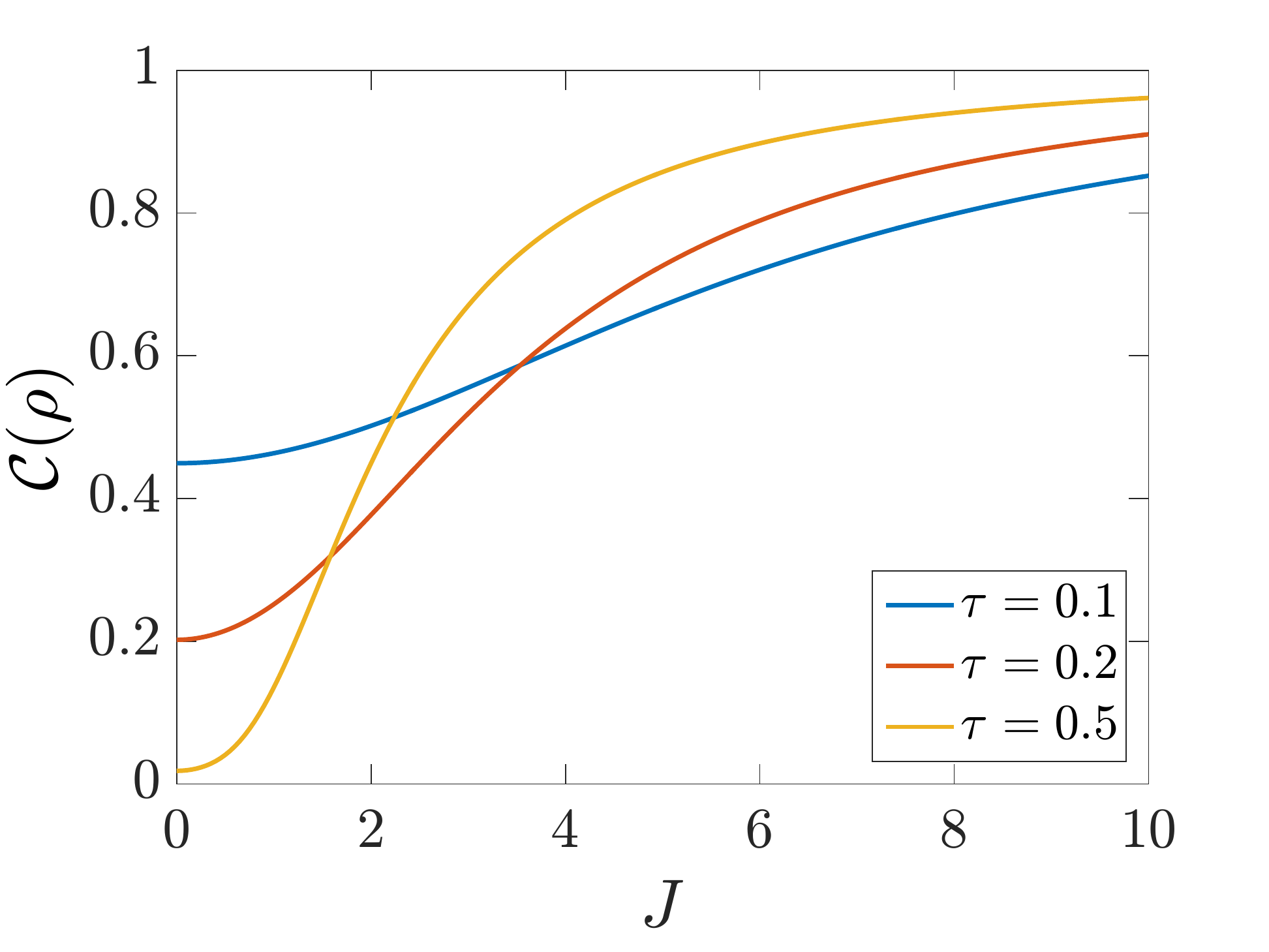}
\end{centering}
\caption{(Color online) Dynamics of entanglement, given by the concurrence $\cl{C}$, as a function of the boson-boson coupling strength, $J$, for a two qubit system evolving via Eq.~\eqref{qq_eff}, for different dissipation timescales $\tau$ (see legend). The two qubits are initialized in the maximally entangled state, $|\psi_0\rangle=\frac{1}{\sqrt{2}}(|00\rangle + |11\rangle)$. The qubit-boson coupling parameters were chosen such that $Tg_A^2=Tg_B^2=1$ arb.~units. This result is valid only in the regime $T \gg \tau$. Time is measured in arbitrary units ($J,g_\alpha$ are inverse time).}
\label{entanglement-fig}
\end{figure}

We also consider the dynamics of the entanglement of the two atoms between modules. The entanglement can be quantified by the concurrence given by $\cl{C}(\rho) = \max(0,\lambda_1-\lambda_3-\lambda_3-\lambda_4)$, where $\lambda_i$ are the eigenvalues in decreasing order of $\sqrt{\sqrt{\rho}\tilde{\rho} \sqrt{\rho}}$ where $\tilde{\rho} = \sigma^y \otimes \sigma^y \rho^\ast \sigma^y \otimes \sigma^y$. 
We consider the effective dynamics for an initially maximally entangled two-qubit system, evolving via Eq.~\eqref{qq_eff}. The effect of this dissipative evolution, of course will result in a degradation of the entanglement. Remarkably however, such degradation can be controlled by increasing the coupling strength $J$. We also find a similar effect, as expected, by increasing $\tau$ (i.e. decreasing the dissipative rate), see Fig.~(\ref{entanglement-fig}). This result is also accurate up to $O(\sqrt{\tau/T})$.

\subsubsection{$z$-dephasing} \label{sect:mixed}
By altering the type of noise acting on the system, we can enact completely different dynamics. In particular, dephasing in the $z$ direction, can be used to prepare mixed states.
\par
Consider two coupled DGMs, each containing $N$ physical qubits, undergoing dephasing in the $z$ direction, that is, with Lindbladian 
\be
\label{eq:z}
\cl{L}_0(\rho)=\tau^{-1}(S^z \rho S^z - \frac{1}{2}\{S^z S^z,\rho\}).
\ee
The SSS contains all states $\Pi_{(J,m)} := |J,m\rangle \langle J,m|$, where $|J,m\rangle \in \cl{H}_J$. Consider the Hamiltonian 
\be
\label{eq:zHam}
K=g(S^+ \otimes S^- + S^-\otimes S^+),
\ee
where the tensor ordering respects the ordering of the two DGMs.
The projector over the SSS is $\cl{P}_0(X)=\sum_{J,m}\Pi_{(J,m)}X\Pi_{(J,m)}$,  (for $m=-J,-J+1,\dots,J$), for quantum states $X\in \text{L}(\cl{H}^{\otimes N})$.
\par
Now, as an example, take a pair of coupled modules, with $N=1$. The steady states in each DGM are of the form $a\Pi_{(1/2,-1/2)} + (1-a)\Pi_{(1/2,1/2)}$, for any $a\in[0,1]$. Consider initializing the system in the (steady) state $\rho_0=|0\rangle \langle 0| \otimes|1\rangle \langle 1|$, where $|0\rangle:=|1/2,-1/2\rangle,|1\rangle:=|1,2,1/2\rangle$. The first non-zero effective term actually enters at second-order (i.e. Eq. \eqref{eq:2ndOrder}), and one can show that the effective dynamics are then given by (see Appendix~\ref{sect:derivation_eqMixed}),
\be
\label{eq:mixed}
\begin{split}
\cl{E}_{eff}(\rho_0)=\frac{1+e^{-2Tg^2\tau}}{2}|0\rangle \langle 0|\otimes |1\rangle \langle 1| \\ + 
\frac{1-e^{-2Tg^2\tau}}{2}|1\rangle \langle 1| \otimes |0\rangle \langle 0|,
\end{split}
\ee
where $\cl{E}_{eff}:=e^{T\cl{L}_{eff}}$, where $Tg^2=O(1)$ (and it is assumed that $T \gg \tau$). In particular, this is the state of the system, up to arbitrarily small errors, after the evolution under Eq.~\eqref{eq:z}, and \eqref{eq:zHam}.
We see that by varying the parameters $g,\tau$ we can transfer, and tune, the populations between two modules, that is, we can dissipatively generate a one dimensional family of correlated two-qubit states. 

\section{Robustness}\label{sect:robustness}
The techniques outlined so far are robust with respect to a certain class of errors. For example, by Eq.~\eqref{eq:dissi}, any perturbation to the control Hamiltonian $K$, say $K\rightarrow K+K'$, results in the same effective  dynamics  provided that $\cl{P}_0\cl{K'P}_0=0$. 
Clearly an analogous result holds also for Lindbladian perturbations with similar properties \cite{zanardi:emergingUnitarity}.

\subsection{Hamiltonian Errors}
We demonstrate in the setting of DGMs, that this robustness holds throughout the network by considering errors to Eq.~\eqref{DGM_ham} of the form
\be
K_{(i,j)} \rightarrow K_{(i,j)}+V_{(i,j)},
\ee
where $V_{(i,j)}$ represents a Hamiltonian encoding error between the modules $i,j$, satisfying $\cl{P}_0\cl{V}_{(i,j)}\cl{P}_0=0$, where $\cl{V}=-i[V,\cdot]$. Collectively then, as these terms only enter the dynamics at most at second order, they give rise to the same effective dynamics. 

Consider, as an illustrative example, two DGMs, each with two qubits collectively dissipating as per Eq.~\eqref{eq:diss}. As discussed in Sect.~\ref{sect:coherent}, each DGM has a two-dimensional DFS = $\text{span}\{|\bar 0 \rangle, |\bar 1 \rangle \}$. We denote the remaining two basis states of the full Hilbert space of each DGM by $|e_0\rangle:=|1,0\rangle,|e_1\rangle:=|1,1\rangle$, using the angular momentum notation, as in Sect.~\ref{sect:coherent}. Errors to the encoding Hamiltonian (e.g. Eq.~\eqref{2dgm_ham}) of the form 
\be
\label{eq:error}
V:=\sum_{\substack{\alpha, \beta = e_0,e_1 \\  i,j=0,1}}\zeta_{\alpha \beta i j} |\alpha \rangle \langle  \bar i | \otimes |\beta \rangle \langle \bar j |+H.c.,
\ee
will be projected out at first order (since $\Pi_{12}V \Pi_{12}=0$, where $\Pi_{12}=\Pi\otimes \Pi$, and $\Pi$ is defined as in Sect.~\ref{sect:coherent}). The tensor ordering respects that of the Hilbert space for the two DGMs. We call $\zeta$ the `error matrix'.

\begin{figure}[h]
\includegraphics[scale=0.4]{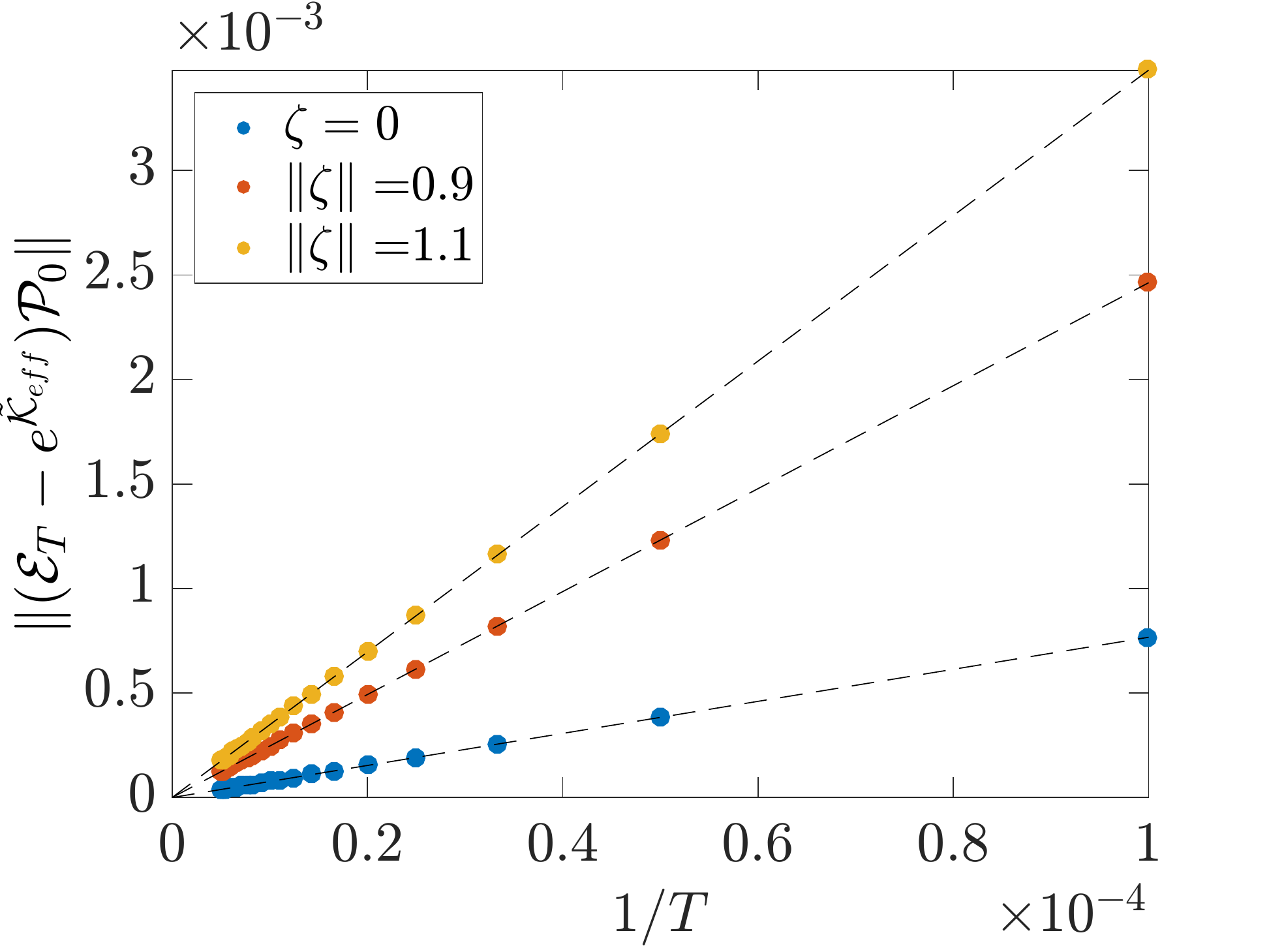}
\caption{(Color online) Robustness to Hamiltonian errors. We consider an unperturbed dissipation given by Eq.~\eqref{eq:diss} plus a control Hamiltonian Eq.~\eqref{2dgm_ham} and an error term of the form Eq.~\eqref{eq:error}. The error matrix $\zeta$ has random complex entries (for two fixed, non-zero magnitudes). We plot the distance between the exact dynamics (i.e.~the dynamics with error terms), to the effective dynamics as governed by Eq.~\eqref{eq:entangling}. Blue (dark gray) data points are without error. The dissipative time-scales are fixed at $\tau_1=\tau_2=1$ arb.~units, and we set $Tg=2$. Norms are calculated using the maximum singular value of the maps realized as matrices. Time is measured in arbitrary units.}
\label{fig:error}
\end{figure}

Note that, because of Eq.~\eqref{eq:dissi}, such Hamiltonian errors give rise to the same effective dynamics accurate up to $O(\tau/T)$. 
We demonstrate the dynamics under such an error in Fig.~\ref{fig:error}. We notice two important properties of this figure. 1) The dynamics with non-zero error matrix $\zeta$ still result in overall error with respect to the effective evolution that is linear in $1/T$. 2) With $\zeta \neq 0$, the overall error, $\|(\cl{E}_T - e^{\cl{\tilde{K}}_{eff}})\cl{P}_0\|$, is strictly greater than the case $\zeta =0$, as expected.

\subsection{Lindbladian Errors}
\begin{figure}[h]
\vspace{1mm}
\includegraphics[scale=0.4]{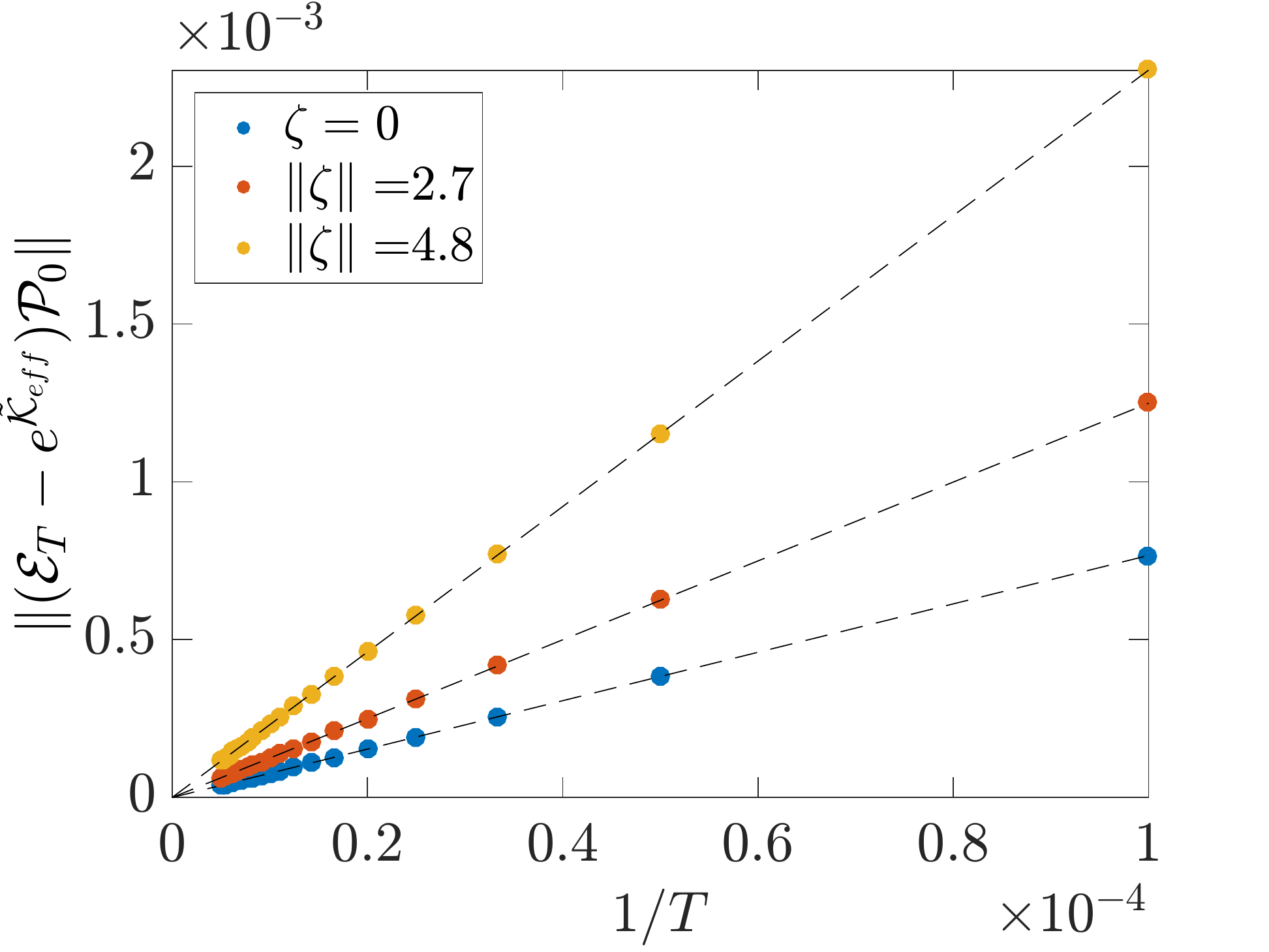}
\caption{(Color online) Robustness to Lindbladian errors. We consider an unperturbed dissipation given by Eq.~\eqref{eq:diss} plus a Lindbladian error term  of the form Eq.~\eqref{eq:L1} with $\eta_i = V$ as in Eq.~\eqref{eq:error}. We show results for  the unperturbed case, $\zeta=0$ (blue/dark gray points), and two non-zero perturbations. The dissipative time-scales are fixed at $\tau_1=\tau_2=1$ arb.~units,  and we set $Tg=2$. Norms are calculated using the maximum singular value of the maps realized as matrices. Time is measured in arbitrary units.}
\label{fig:lindbladError}
\end{figure}

This analysis can be extended to include the case where there are errors to the Lindblad operators, $L_i$, defining a Lindbladian of the form
\be
\cl{L}_0(\rho) = \sum_{i}( L_i \rho L_i^\dagger -\frac{1}{2}\{L_i ^\dagger L_i , \rho\}). \label{eq:L0}
\ee
Consider an error of the form $L_i \rightarrow L_i + \eta_i$, where the $\eta_i$ are assumed to be $O(1/T)$. To $O(1/T)$ then, we have $\cl{L}_0\rightarrow \cl{L}_0+\cl{L}_1$, where
\be
\cl{L}_1(\rho) = \sum_i (\eta_i\rho L_i^\dagger -\frac{1}{2}\{L_i^\dagger \eta_i,\rho\}) + H.c.\label{eq:L1}
\ee
As in the previous subsection, 
all perturbations of this form such that $\cl{P}_0\cl{L}_1\cl{P}_0=0$ give rise to the same effective dynamics.
We illustrate this by following the same example in the previous subsection with $\cl{L}_0$ given by Eq.~\eqref{eq:diss}. It is easy to verify that by choosing $\eta_i = V$ in Eq.~\eqref{eq:L1} [$V$ as in Eq.~\eqref{eq:error}] one obtains $\cl{P}_0\cl{L}_1\cl{P}_0=0$ \footnote{From Sect.~\ref{sect:coherent}, states in the DFS are linear combinations of terms of the form \unexpanded{$\chi_{\text{ss}}:=| \bar n_0 \rangle \langle \bar n_1 | \otimes | \bar n_2 \rangle \langle \bar n_3 |$}, \unexpanded{$n_i \in \{0,1\}$}. There is a single Lindblad operator for each DGM, $S^-$. Non-zero terms of \unexpanded{$\cl{L}_1 (\chi_\text{ss})$}, are of the form \unexpanded{$X\otimes |e_i \rangle \langle \bar n_j |$}, or \unexpanded{$|e_i \rangle \langle \bar n_j | \otimes X$}, for some $X$  (or Hermitian conjugate). This is sufficient to see that \unexpanded{$\cl{P}_0\cl{L}_1(\chi_\text{ss})=0$} (regardless of $X$), hence \unexpanded{$\cl{P}_0\cl{L}_1\cl{P}_0=0$}}. We numerically validate this in Fig.~\ref{fig:lindbladError}, where we do indeed see extra $O(1/T)$ errors resulting from the Lindbladian perturbations. 

\section{Conclusion}
It is nowadays clear that quantum dissipation and decoherence can be turned into resources for the design and implementation of quantum information processing tasks.
In this paper, building upon the ideas and findings of Refs.  \cite{zanardi-dissipation-2014,dissi_2nd_order:Zanardi2016}, we  discussed a scalable  architecture consisting of a network of  dissipation-generated modules (DGMs). The basic  modules comprise a few qubits 
undergoing an internal fast dissipative process. Different modules  are  then coherently connected to one another via Hamiltonian couplings. 

We have shown that a dissipation-assisted modular network of this type can be used to perform a diverse number of tasks, such as enacting any single qubit gate, preparing entangled states, and can even be used to perform universal information processing tasks. 

We gave the explicit construction of a two-module network, which can be used to preserve the coherence of a single qubit, and  the entanglement between two qubits in separate, lossy cavities. Finally, we demonstrated that under certain Hamiltonian -- and even Lindbladian --  perturbations, the dynamics is  unaffected, up to errors vanishing in the fast dissipation limit.

This inbuilt robustness seems to indicate the potential practical relevance of the  architecture considered in this paper.
Conceptually, the scalability and computational universality of our dissipation-assisted modular networks show yet another  way in which
dissipation can be turned into a powerful resource for quantum manipulations. 

Acknowledgements.- This work was partially supported
by the ARO MURI grant  W911NF-11-1-0268 and W911NF-15-1-0582.
\bibliography{refs}

\begin{thebibliography}{28}%
\makeatletter
\providecommand \@ifxundefined [1]{%
 \@ifx{#1\undefined}
}%
\providecommand \@ifnum [1]{%
 \ifnum #1\expandafter \@firstoftwo
 \else \expandafter \@secondoftwo
 \fi
}%
\providecommand \@ifx [1]{%
 \ifx #1\expandafter \@firstoftwo
 \else \expandafter \@secondoftwo
 \fi
}%
\providecommand \natexlab [1]{#1}%
\providecommand \enquote  [1]{``#1''}%
\providecommand \bibnamefont  [1]{#1}%
\providecommand \bibfnamefont [1]{#1}%
\providecommand \citenamefont [1]{#1}%
\providecommand \href@noop [0]{\@secondoftwo}%
\providecommand \href [0]{\begingroup \@sanitize@url \@href}%
\providecommand \@href[1]{\@@startlink{#1}\@@href}%
\providecommand \@@href[1]{\endgroup#1\@@endlink}%
\providecommand \@sanitize@url [0]{\catcode `\\12\catcode `\$12\catcode
  `\&12\catcode `\#12\catcode `\^12\catcode `\_12\catcode `\%12\relax}%
\providecommand \@@startlink[1]{}%
\providecommand \@@endlink[0]{}%
\providecommand \url  [0]{\begingroup\@sanitize@url \@url }%
\providecommand \@url [1]{\endgroup\@href {#1}{\urlprefix }}%
\providecommand \urlprefix  [0]{URL }%
\providecommand \Eprint [0]{\href }%
\providecommand \doibase [0]{http://dx.doi.org/}%
\providecommand \selectlanguage [0]{\@gobble}%
\providecommand \bibinfo  [0]{\@secondoftwo}%
\providecommand \bibfield  [0]{\@secondoftwo}%
\providecommand \translation [1]{[#1]}%
\providecommand \BibitemOpen [0]{}%
\providecommand \bibitemStop [0]{}%
\providecommand \bibitemNoStop [0]{.\EOS\space}%
\providecommand \EOS [0]{\spacefactor3000\relax}%
\providecommand \BibitemShut  [1]{\csname bibitem#1\endcsname}%
\let\auto@bib@innerbib\@empty
\bibitem [{\citenamefont {Lidar}\ and\ \citenamefont
  {Brun}(2013)}]{Lidar-Brun:book}%
  \BibitemOpen
  \bibinfo {editor} {\bibfnamefont {D.A.}\ \bibnamefont {Lidar}}\ and\ \bibinfo
  {editor} {\bibfnamefont {T.A.}\ \bibnamefont {Brun}},\ eds.,\ \href
  {http://www.cambridge.org/9780521897877} {\emph {\bibinfo {title} {Quantum
  Error Correction}}}\ (\bibinfo  {publisher} {Cambridge University Press},\
  \bibinfo {address} {{Cambridge, UK}},\ \bibinfo {year} {2013})\BibitemShut
  {NoStop}%
\bibitem [{\citenamefont {Gottesman}(1996)}]{Gottesman:1996fk}%
  \BibitemOpen
  \bibfield  {author} {\bibinfo {author} {\bibfnamefont {D.}~\bibnamefont
  {Gottesman}},\ }\bibfield  {title} {\enquote {\bibinfo {title} {Class of
  quantum error-correcting codes saturating the quantum hamming bound},}\
  }\href {\doibase 10.1103/PhysRevA.54.1862} {\bibfield  {journal} {\bibinfo
  {journal} {{Phys. Rev. A}}\ }\textbf {\bibinfo {volume} {54}},\ \bibinfo
  {pages} {1862} (\bibinfo {year} {1996})}\BibitemShut {NoStop}%
\bibitem [{\citenamefont {{D. Kielpinski, V. Meyer, M.A. Rowe, C.A. Sackett,
  W.M. Itano, C. Monroe, and D.J. Wineland}}(2001)}]{Kielpinski:01}%
  \BibitemOpen
  \bibfield  {author} {\bibinfo {author} {\bibnamefont {{D. Kielpinski, V.
  Meyer, M.A. Rowe, C.A. Sackett, W.M. Itano, C. Monroe, and D.J. Wineland}}},\
  }\bibfield  {title} {\enquote {\bibinfo {title} {{A Decoherence-Free Quantum
  Memory Using Trapped Ions}},}\ }\href@noop {} {\bibfield  {journal} {\bibinfo
   {journal} {Science}\ }\textbf {\bibinfo {volume} {291}},\ \bibinfo {pages}
  {1013} (\bibinfo {year} {2001})}\BibitemShut {NoStop}%
\bibitem [{\citenamefont {Zanardi}\ and\ \citenamefont
  {Rasetti}(1997)}]{Zanardi:97c}%
  \BibitemOpen
  \bibfield  {author} {\bibinfo {author} {\bibfnamefont {P.}~\bibnamefont
  {Zanardi}}\ and\ \bibinfo {author} {\bibfnamefont {M.}~\bibnamefont
  {Rasetti}},\ }\bibfield  {title} {\enquote {\bibinfo {title} {Noiseless
  quantum codes},}\ }\href
  {http://link.aps.org/doi/10.1103/PhysRevLett.79.3306} {\bibfield  {journal}
  {\bibinfo  {journal} {Physical Review Letters}\ }\textbf {\bibinfo {volume}
  {79}},\ \bibinfo {pages} {3306--3309} (\bibinfo {year} {1997})}\BibitemShut
  {NoStop}%
\bibitem [{\citenamefont {Lidar}\ \emph {et~al.}(1998)\citenamefont {Lidar},
  \citenamefont {Chuang},\ and\ \citenamefont {Whaley}}]{Lidar:1998fk}%
  \BibitemOpen
  \bibfield  {author} {\bibinfo {author} {\bibfnamefont {D.~A.}\ \bibnamefont
  {Lidar}}, \bibinfo {author} {\bibfnamefont {I.~L.}\ \bibnamefont {Chuang}}, \
  and\ \bibinfo {author} {\bibfnamefont {K.~B.}\ \bibnamefont {Whaley}},\
  }\bibfield  {title} {\enquote {\bibinfo {title} {Decoherence-free subspaces
  for quantum computation},}\ }\href
  {http://link.aps.org/doi/10.1103/PhysRevLett.81.2594} {\bibfield  {journal}
  {\bibinfo  {journal} {Phys. Rev. Lett.}\ }\textbf {\bibinfo {volume} {81}},\
  \bibinfo {pages} {2594--2597} (\bibinfo {year} {1998})}\BibitemShut {NoStop}%
\bibitem [{\citenamefont {Knill}\ \emph {et~al.}(2000)\citenamefont {Knill},
  \citenamefont {Laflamme},\ and\ \citenamefont {Viola}}]{Knill:2000dq}%
  \BibitemOpen
  \bibfield  {author} {\bibinfo {author} {\bibfnamefont {Emanuel}\ \bibnamefont
  {Knill}}, \bibinfo {author} {\bibfnamefont {Raymond}\ \bibnamefont
  {Laflamme}}, \ and\ \bibinfo {author} {\bibfnamefont {Lorenza}\ \bibnamefont
  {Viola}},\ }\bibfield  {title} {\enquote {\bibinfo {title} {Theory of quantum
  error correction for general noise},}\ }\href
  {http://link.aps.org/doi/10.1103/PhysRevLett.84.2525} {\bibfield  {journal}
  {\bibinfo  {journal} {{Phys.~Rev.~Lett.}}\ }\textbf {\bibinfo {volume}
  {84}},\ \bibinfo {pages} {2525--2528} (\bibinfo {year} {2000})}\BibitemShut
  {NoStop}%
\bibitem [{\citenamefont {Zanardi}(2000)}]{Zanardi:99d}%
  \BibitemOpen
  \bibfield  {author} {\bibinfo {author} {\bibfnamefont {P.}~\bibnamefont
  {Zanardi}},\ }\bibfield  {title} {\enquote {\bibinfo {title} {Stabilizing
  quantum information},}\ }\href@noop {} {\bibfield  {journal} {\bibinfo
  {journal} {Phys. Rev. A}\ }\textbf {\bibinfo {volume} {63}},\ \bibinfo
  {pages} {012301} (\bibinfo {year} {2000})}\BibitemShut {NoStop}%
\bibitem [{\citenamefont {Zanardi}\ and\ \citenamefont
  {Lloyd}(2003)}]{Zanardi:2003c}%
  \BibitemOpen
  \bibfield  {author} {\bibinfo {author} {\bibfnamefont {Paolo}\ \bibnamefont
  {Zanardi}}\ and\ \bibinfo {author} {\bibfnamefont {Seth}\ \bibnamefont
  {Lloyd}},\ }\bibfield  {title} {\enquote {\bibinfo {title} {Topological
  protection and quantum noiseless subsystems},}\ }\href
  {http://link.aps.org/doi/10.1103/PhysRevLett.90.067902} {\bibfield  {journal}
  {\bibinfo  {journal} {Physical Review Letters}\ }\textbf {\bibinfo {volume}
  {90}},\ \bibinfo {pages} {067902} (\bibinfo {year} {2003})}\BibitemShut
  {NoStop}%
\bibitem [{\citenamefont {Kraus}\ \emph {et~al.}(2008)\citenamefont {Kraus},
  \citenamefont {B{\"u}chler}, \citenamefont {Diehl}, \citenamefont {Kantian},
  \citenamefont {Micheli},\ and\ \citenamefont {Zoller}}]{Kraus-prep}%
  \BibitemOpen
  \bibfield  {author} {\bibinfo {author} {\bibfnamefont {B.}~\bibnamefont
  {Kraus}}, \bibinfo {author} {\bibfnamefont {H.~P.}\ \bibnamefont
  {B{\"u}chler}}, \bibinfo {author} {\bibfnamefont {S.}~\bibnamefont {Diehl}},
  \bibinfo {author} {\bibfnamefont {A.}~\bibnamefont {Kantian}}, \bibinfo
  {author} {\bibfnamefont {A.}~\bibnamefont {Micheli}}, \ and\ \bibinfo
  {author} {\bibfnamefont {P.}~\bibnamefont {Zoller}},\ }\bibfield  {title}
  {\enquote {\bibinfo {title} {Preparation of entangled states by quantum
  markov processes},}\ }\href
  {http://link.aps.org/doi/10.1103/PhysRevA.78.042307} {\bibfield  {journal}
  {\bibinfo  {journal} {Physical Review A}\ }\textbf {\bibinfo {volume} {78}},\
  \bibinfo {pages} {042307} (\bibinfo {year} {2008})}\BibitemShut {NoStop}%
\bibitem [{\citenamefont {Barreiro}\ \emph {et~al.}(2011)\citenamefont
  {Barreiro}, \citenamefont {Muller}, \citenamefont {Schindler}, \citenamefont
  {Nigg}, \citenamefont {Monz}, \citenamefont {Chwalla}, \citenamefont
  {Hennrich}, \citenamefont {Roos}, \citenamefont {Zoller},\ and\ \citenamefont
  {Blatt}}]{barreiro2011open}%
  \BibitemOpen
  \bibfield  {author} {\bibinfo {author} {\bibfnamefont {Julio~T.}\
  \bibnamefont {Barreiro}}, \bibinfo {author} {\bibfnamefont {Markus}\
  \bibnamefont {Muller}}, \bibinfo {author} {\bibfnamefont {Philipp}\
  \bibnamefont {Schindler}}, \bibinfo {author} {\bibfnamefont {Daniel}\
  \bibnamefont {Nigg}}, \bibinfo {author} {\bibfnamefont {Thomas}\ \bibnamefont
  {Monz}}, \bibinfo {author} {\bibfnamefont {Michael}\ \bibnamefont {Chwalla}},
  \bibinfo {author} {\bibfnamefont {Markus}\ \bibnamefont {Hennrich}}, \bibinfo
  {author} {\bibfnamefont {Christian~F.}\ \bibnamefont {Roos}}, \bibinfo
  {author} {\bibfnamefont {Peter}\ \bibnamefont {Zoller}}, \ and\ \bibinfo
  {author} {\bibfnamefont {Rainer}\ \bibnamefont {Blatt}},\ }\bibfield  {title}
  {\enquote {\bibinfo {title} {An open-system quantum simulator with trapped
  ions},}\ }\href {http://dx.doi.org/10.1038/nature09801} {\bibfield  {journal}
  {\bibinfo  {journal} {Nature}\ }\textbf {\bibinfo {volume} {470}},\ \bibinfo
  {pages} {486--491} (\bibinfo {year} {2011})}\BibitemShut {NoStop}%
\bibitem [{\citenamefont {Childs}\ \emph {et~al.}(2001)\citenamefont {Childs},
  \citenamefont {Farhi},\ and\ \citenamefont {Preskill}}]{childs:01}%
  \BibitemOpen
  \bibfield  {author} {\bibinfo {author} {\bibfnamefont {Andrew~M.}\
  \bibnamefont {Childs}}, \bibinfo {author} {\bibfnamefont {Edward}\
  \bibnamefont {Farhi}}, \ and\ \bibinfo {author} {\bibfnamefont {John}\
  \bibnamefont {Preskill}},\ }\bibfield  {title} {\enquote {\bibinfo {title}
  {Robustness of adiabatic quantum computation},}\ }\href@noop {} {\bibfield
  {journal} {\bibinfo  {journal} {Phys. Rev. A}\ }\textbf {\bibinfo {volume}
  {65}},\ \bibinfo {pages} {012322} (\bibinfo {year} {2001})}\BibitemShut
  {NoStop}%
\bibitem [{\citenamefont {Sarandy}\ and\ \citenamefont
  {Lidar}(2005)}]{sarandy:05}%
  \BibitemOpen
  \bibfield  {author} {\bibinfo {author} {\bibfnamefont {M.~S.}\ \bibnamefont
  {Sarandy}}\ and\ \bibinfo {author} {\bibfnamefont {D.~A.}\ \bibnamefont
  {Lidar}},\ }\bibfield  {title} {\enquote {\bibinfo {title} {Adiabatic quantum
  computation in open systems},}\ }\href@noop {} {\bibfield  {journal}
  {\bibinfo  {journal} {Phys. Rev. Lett.}\ }\textbf {\bibinfo {volume} {95}},\
  \bibinfo {pages} {250503} (\bibinfo {year} {2005})}\BibitemShut {NoStop}%
\bibitem [{\citenamefont {Zanardi}\ and\ \citenamefont
  {Venuti}(2014)}]{zanardi-dissipation-2014}%
  \BibitemOpen
  \bibfield  {author} {\bibinfo {author} {\bibfnamefont {Paolo}\ \bibnamefont
  {Zanardi}}\ and\ \bibinfo {author} {\bibfnamefont {Lorenzo~Campos}\
  \bibnamefont {Venuti}},\ }\bibfield  {title} {\enquote {\bibinfo {title}
  {Coherent quantum dynamics in steady-state manifolds of strongly dissipative
  systems},}\ }\href
  {http://journals.aps.org/prl/abstract/10.1103/PhysRevLett.113.240406}
  {\bibfield  {journal} {\bibinfo  {journal} {{Phys.~Rev.~Lett.}}\ }\textbf
  {\bibinfo {volume} {113}},\ \bibinfo {pages} {240406} (\bibinfo {year}
  {2014})}\BibitemShut {NoStop}%
\bibitem [{\citenamefont {{Paolo Zanardi, Jeffrey Marshall, and Lorenzo Campos
  Venuti}}(2016)}]{dissi_2nd_order:Zanardi2016}%
  \BibitemOpen
  \bibfield  {author} {\bibinfo {author} {\bibnamefont {{Paolo Zanardi, Jeffrey
  Marshall, and Lorenzo Campos Venuti}}},\ }\bibfield  {title} {\enquote
  {\bibinfo {title} {Dissipative universal {L}indbladian simulation},}\
  }\href@noop {} {\bibfield  {journal} {\bibinfo  {journal} {{Phys. Rev. A}}\
  }\textbf {\bibinfo {volume} {93}},\ \bibinfo {pages} {022312} (\bibinfo
  {year} {2016})}\BibitemShut {NoStop}%
\bibitem [{\citenamefont {Sato}(1977)}]{sato_ergodic_1977}%
  \BibitemOpen
  \bibfield  {author} {\bibinfo {author} {\bibfnamefont {Ryotaro}\ \bibnamefont
  {Sato}},\ }\bibfield  {title} {\enquote {\bibinfo {title} {Ergodic theorems
  for semigroups of positive operators},}\ }\href {\doibase
  10.2969/jmsj/02940591} {\bibfield  {journal} {\bibinfo  {journal} {J. Math.
  Soc. Japan}\ }\textbf {\bibinfo {volume} {29}},\ \bibinfo {pages} {591--606}
  (\bibinfo {year} {1977})}\BibitemShut {NoStop}%
\bibitem [{\citenamefont {Venuti}\ \emph {et~al.}(2016)\citenamefont {Venuti},
  \citenamefont {Albash}, \citenamefont {Lidar},\ and\ \citenamefont
  {Zanardi}}]{venuti_adiabaticity_2016}%
  \BibitemOpen
  \bibfield  {author} {\bibinfo {author} {\bibfnamefont {Lorenzo~Campos}\
  \bibnamefont {Venuti}}, \bibinfo {author} {\bibfnamefont {Tameem}\
  \bibnamefont {Albash}}, \bibinfo {author} {\bibfnamefont {Daniel~A.}\
  \bibnamefont {Lidar}}, \ and\ \bibinfo {author} {\bibfnamefont {Paolo}\
  \bibnamefont {Zanardi}},\ }\bibfield  {title} {\enquote {\bibinfo {title}
  {Adiabaticity in open quantum systems},}\ }\href
  {http://link.aps.org/doi/10.1103/PhysRevA.93.032118} {\bibfield  {journal}
  {\bibinfo  {journal} {Phys. Rev. A}\ }\textbf {\bibinfo {volume} {93}},\
  \bibinfo {pages} {032118} (\bibinfo {year} {2016})}\BibitemShut {NoStop}%
\bibitem [{Note1()}]{Note1}%
  \BibitemOpen
  \bibinfo {note} {This estimate is a worst case scenario which assumes that
  all the couplings $K_{(i,j)} $are always on}\BibitemShut {NoStop}%
\bibitem [{\citenamefont {{Victor V. Albert, Barry Bradlyn, Martin Fraas, and
  Liang Jiang}}(2015)}]{Albert:geometry_response}%
  \BibitemOpen
  \bibfield  {author} {\bibinfo {author} {\bibnamefont {{Victor V. Albert,
  Barry Bradlyn, Martin Fraas, and Liang Jiang}}},\ }\bibfield  {title}
  {\enquote {\bibinfo {title} {Geometry and response of {L}indbladians},}\
  }\href@noop {} {\bibfield  {journal} {\bibinfo  {journal}
  {{arXiv:1512.08079}}\ } (\bibinfo {year} {2015})}\BibitemShut {NoStop}%
\bibitem [{Note2()}]{Note2}%
  \BibitemOpen
  \bibinfo {note} {$|0,0\rangle =\frac {1}{\sqrt {2}}(|01\rangle -|10\rangle
  ),|1,-1\rangle =|00\rangle $.}\BibitemShut {Stop}%
\bibitem [{Note3()}]{Note3}%
  \BibitemOpen
  \bibinfo {note} {Consider the Hamiltonian $K=\sigma ^z \otimes \sigma ^z$.
  Clearly the action of this on $|\bar 1\rangle $ is the identity, whereas
  results in a minus sign on $|\bar 0\rangle $. Therefore, this will result in
  a logical $\sigma ^z$ Hamiltonian. $\bar \sigma ^x$ is derived in a similar
  manner.}\BibitemShut {Stop}%
\bibitem [{\citenamefont {{R.C. Bialczak, M. Ansmann, M. Hofheinz, E. Lucero,
  M. Neeley, A.D. O'Connell, D. Sank, H. Wang, J. Wenner, M. Steffen, A. N.
  Cleland, J. M. Martinis}}(2010)}]{bialczak:q_tomography}%
  \BibitemOpen
  \bibfield  {author} {\bibinfo {author} {\bibnamefont {{R.C. Bialczak, M.
  Ansmann, M. Hofheinz, E. Lucero, M. Neeley, A.D. O'Connell, D. Sank, H. Wang,
  J. Wenner, M. Steffen, A. N. Cleland, J. M. Martinis}}},\ }\bibfield  {title}
  {\enquote {\bibinfo {title} {{Quantum process tomography of a universal
  entangling gate implemented with Josephson phase qubits}},}\ }\href@noop {}
  {\bibfield  {journal} {\bibinfo  {journal} {Nature Physics}\ }\textbf
  {\bibinfo {volume} {6}},\ \bibinfo {pages} {409--413} (\bibinfo {year}
  {2010})}\BibitemShut {NoStop}%
\bibitem [{Note4()}]{Note4}%
  \BibitemOpen
  \bibinfo {note} {We define $|\bar 0\protect \rangle := |1/2,0,-1/2\protect
  \rangle ,\, |\bar 1\protect \rangle :=|1/2,1,-1/2\protect \rangle ,\, |\bar 2
  \protect \rangle := |3/2,-3/2 \protect \rangle $, where the second index on
  the first two states indicates which of the two spin 1/2 sectors it belongs.
  Qubit notation: $|1/2,0,-1/2\protect \rangle = \frac {1}{\sqrt
  {2}}(|010\protect \rangle - |100\protect \rangle ),\, |1/2,1,-1/2\protect
  \rangle =\frac {1}{\sqrt {6}}(|100\protect \rangle + |100\protect \rangle
  -2|001\protect \rangle ),\, |3/2,-3/2\protect \rangle =|000\protect \rangle
  $.}\BibitemShut {Stop}%
\bibitem [{\citenamefont {Sete}\ \emph {et~al.}(2015)\citenamefont {Sete},
  \citenamefont {Martinis},\ and\ \citenamefont
  {Korotkov}}]{sete_quantum_2015}%
  \BibitemOpen
  \bibfield  {author} {\bibinfo {author} {\bibfnamefont {Eyob~A.}\ \bibnamefont
  {Sete}}, \bibinfo {author} {\bibfnamefont {John~M.}\ \bibnamefont
  {Martinis}}, \ and\ \bibinfo {author} {\bibfnamefont {Alexander~N.}\
  \bibnamefont {Korotkov}},\ }\bibfield  {title} {\enquote {\bibinfo {title}
  {Quantum theory of a bandpass {Purcell} filter for qubit readout},}\
  }\href@noop {} {\bibfield  {journal} {\bibinfo  {journal} {Phys. Rev. A}\
  }\textbf {\bibinfo {volume} {92}},\ \bibinfo {pages} {012325} (\bibinfo
  {year} {2015})}\BibitemShut {NoStop}%
\bibitem [{\citenamefont {{Zhong-Xiao Man, Yun-Jie Xia, and Rosario Lo
  Franco}}(2015)}]{man:2015_cavity}%
  \BibitemOpen
  \bibfield  {author} {\bibinfo {author} {\bibnamefont {{Zhong-Xiao Man,
  Yun-Jie Xia, and Rosario Lo Franco}}},\ }\bibfield  {title} {\enquote
  {\bibinfo {title} {{Cavity-based architecture to preserve quantum coherence
  and entanglement}},}\ }\href@noop {} {\bibfield  {journal} {\bibinfo
  {journal} {Scientific Reports}\ }\textbf {\bibinfo {volume} {5}} (\bibinfo
  {year} {2015})}\BibitemShut {NoStop}%
\bibitem [{\citenamefont {Gonz{\'a}lez-Guti{\'e}rrez}\ \emph
  {et~al.}(2016)\citenamefont {Gonz{\'a}lez-Guti{\'e}rrez}, \citenamefont
  {Villase{\~n}or}, \citenamefont {Pineda},\ and\ \citenamefont
  {Seligman}}]{gonzalez_coherence}%
  \BibitemOpen
  \bibfield  {author} {\bibinfo {author} {\bibfnamefont {C}~\bibnamefont
  {Gonz{\'a}lez-Guti{\'e}rrez}}, \bibinfo {author} {\bibfnamefont
  {E}~\bibnamefont {Villase{\~n}or}}, \bibinfo {author} {\bibfnamefont
  {C}~\bibnamefont {Pineda}}, \ and\ \bibinfo {author} {\bibfnamefont {T~H}\
  \bibnamefont {Seligman}},\ }\bibfield  {title} {\enquote {\bibinfo {title}
  {Stabilizing coherence with nested environments: a numerical study using
  kicked ising models},}\ }\href@noop {} {\bibfield  {journal} {\bibinfo
  {journal} {Physica Scripta}\ }\textbf {\bibinfo {volume} {91}},\ \bibinfo
  {pages} {083001} (\bibinfo {year} {2016})}\BibitemShut {NoStop}%
\bibitem [{Note5()}]{Note5}%
  \BibitemOpen
  \bibinfo {note} {One can calculate the functional form of the coherence, in
  the computational basis ($|i\rangle $, $i=0,1$), by noting that $\cl
  {L}_{eff}^\alpha (|i\rangle _\alpha \langle j |) = -\frac {\tau _{eff,\alpha
  }^{-1}}{2}|i\rangle _\alpha \langle j|$, for $i\not =j$, and $\cl
  {L}_{eff}^\alpha (|1\rangle _\alpha \langle 1|) = \tau _{eff,\alpha
  }^{-1}(|0\rangle _\alpha 0| - |1 \rangle _\alpha \langle 1|)$. Note,
  $|0\rangle _\alpha \langle 0|$ is a steady state. The subscript $\alpha $ on
  the states just indicate the system $\alpha =A,B$.}\BibitemShut {Stop}%
\bibitem [{\citenamefont {{Paolo Zanardi, and Lorenzo Campos
  Venuti}}(2015)}]{zanardi:emergingUnitarity}%
  \BibitemOpen
  \bibfield  {author} {\bibinfo {author} {\bibnamefont {{Paolo Zanardi, and
  Lorenzo Campos Venuti}}},\ }\bibfield  {title} {\enquote {\bibinfo {title}
  {Geometry, robustness, and emerging unitarity in dissipation-projected
  dynamics},}\ }\href@noop {} {\bibfield  {journal} {\bibinfo  {journal}
  {{Phys. Rev. A}}\ }\textbf {\bibinfo {volume} {91}},\ \bibinfo {pages}
  {052324} (\bibinfo {year} {2015})}\BibitemShut {NoStop}%
\bibitem [{Note6()}]{Note6}%
  \BibitemOpen
  \bibinfo {note} {From Sect.~\ref {sect:coherent}, states in the DFS are
  linear combinations of terms of the form $\chi _{\text {ss}}:=| \bar n_0
  \rangle \langle \bar n_1 | \otimes | \bar n_2 \rangle \langle \bar n_3 |$,
  $n_i \in \{0,1\}$. There is a single Lindblad operator for each DGM, $S^-$.
  Non-zero terms of $\cl {L}_1 (\chi _\text {ss})$, are of the form $X\otimes
  |e_i \rangle \langle \bar n_j |$, or $|e_i \rangle \langle \bar n_j | \otimes
  X$, for some $X$ (or Hermitian conjugate). This is sufficient to see that
  $\cl {P}_0\cl {L}_1(\chi _\text {ss})=0$ (regardless of $X$), hence $\cl
  {P}_0\cl {L}_1\cl {P}_0=0$}\BibitemShut {NoStop}%
\end{thebibliography}%

\newpage

\appendix

\section{Dissipative preparation of the CNOT gate} 
 \label{app:cnot}
 
 In Sect.~\ref{sect:coherent} it was shown possible to generate an entangling Hamiltonian of the form Eq.~\eqref{eq:entangling}. Denote this Hamiltonian $K_{\text{swap}}(g)$, where the parameter $g$ is the strength of the Hamiltonian, as given in Eq.~\eqref{eq:entangling}. In the (ordered) basis $\{ |\bar 0 \bar 0 \rangle, |\bar 0 \bar 1 \rangle , |\bar 1 \bar 0 \rangle, |\bar 1 \bar 1 \rangle \}$, where $|\bar 0 \rangle, |\bar 1 \rangle$ are defined in the main text, it is clear that (effective) evolution under $K_{\text{swap}}(\frac{\pi}{2})$, generates the SQ\textit{i}SW$=:S_i$ gate:
 \be
S_i = 
 \left( \begin{array}{cccc}
1 & 0 & 0 & 0 \\
0 & \frac{1}{\sqrt{2}} & -\frac{i}{\sqrt{2}} & 0 \\
0 & -\frac{i}{\sqrt{2}} & \frac{1}{\sqrt{2}} & 0 \\
0 & 0 & 0 & 1 \end{array} \right).
 \ee
 We define qubit rotations by $\theta$ about $x,y$ respectively as $X_{\theta}=e^{-i\frac{\theta}{2}\bar \sigma^x}, Y_{\theta}=e^{-i\frac{\theta}{2}\bar \sigma^y}$. In Sect.~\ref{sect:coherent} it was shown how to generate (effective) logical Pauli Hamiltonians, and thus any single qubit gate. This ability to generate SQ\textit{i}SW, $X_{\theta}$, and $Y_{\theta}$ enables us to create an effective CNOT gate \cite{bialczak:q_tomography}. In particular, 
 \be
 \label{eq:cnot}
 \text{CNOT} = Y_{\frac{-\pi}{2}}^{(1)}X^{(2)}_{\frac{-\pi}{2}}X^{(1)}_{\frac{\pi}{2}}S_iX^{(1)}_{\pi}S_iY^{(1)}_{\frac{\pi}{2}},
 \ee
where the superscript index indicates which DGM we are operating on. To simulate, for example, $X^{(1)}_\pi$, we evolve the two modules (each containing 2 qubits), which are dissipating via Eq.~\eqref{eq:diss}, with Hamiltonian $K=\frac{\pi}{\sqrt{2}}\sigma_2^x \otimes \mathbb{I}$ (see Sect.~\ref{sect:coherent} for reference). As is requisite for this technique, we assume $K$ is scaled by a factor of $1/T$. Joining together several operations of this type (as given by Eq.~\eqref{eq:cnot}), we can in fact construct the CNOT gate, assuming the leakage out of the SSS is negligible (or at least, controllable) at each step. Since the errors at each step are always $O(1/T)$, we expect the scaling, for large $T$, to in fact still be linear in $1/T$.
\par
\begin{figure}[h]
\vspace{1mm}
\begin{centering}
\includegraphics[scale=0.4]{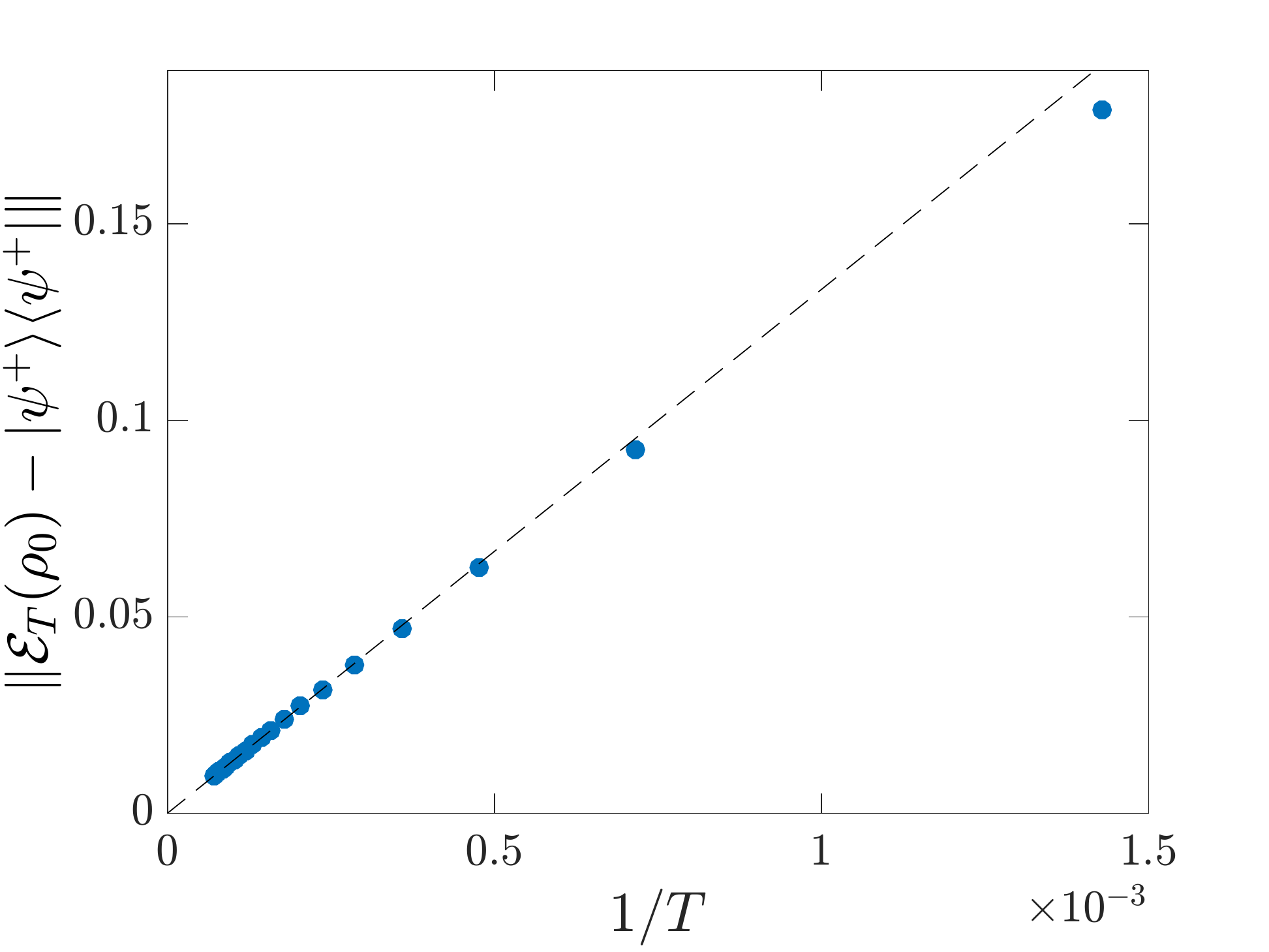}
\end{centering}
\caption{(Color online) Distance between exact evolution, and the target maximally entangled Bell state, $|\psi^+\rangle$ (defined in main text). We initialize the system in the product state $\rho_0=|\psi_0 \rangle \langle \psi_0|$, where $|\psi_0\rangle= \frac{|\bar 0\rangle + |\bar 1\rangle}{\sqrt{2}}|\bar 0\rangle$. The exact evolution $\cl{E}_T=\cl{E}^{(7)}_{T/7}\cl{E}^{(6)}_{T/7}\dots\cl{E}^{(1)}_{T/7}$, where each $\cl{E}^{(i)}_{T/7}$ generates one of the 7 gates of Eq.~\eqref{eq:cnot}, as described in the main text. The dissipative time-scales for the two DGMs were set to $\tau_1=\tau_2=1$ arb. units for this simulation. The norm is the maximum singular value of the maps realized as matrices. The linear fit is on the 10 largest $T$ values. Time is measured in arbitrary units.}
\label{cnot-fig}
\end{figure}

To show our scheme is effective at preparing such a gate, and hence entangled states, we illustrate the ability to prepare the maximally entangled state $|\psi^+\rangle:=\frac{1}{\sqrt{2}}(|\bar 0 \bar 0 \rangle + |\bar 1 \bar 1 \rangle)$, from an initial product state, with arbitrarily small error, in Fig.~\ref{cnot-fig}. We see for larger and larger $T$, the error bound is better approximated as a linear function in $1/T$, as expected.
\par
Fig.~\ref{cnot-fig} shows that despite the 7 separate evolutions required to prepare the CNOT gate, where errors will accumulate at each step, by tuning the total evolution time, $T$, one can arbitrarily control the overall error in the system. Given this ability to perform CNOT and the single qubit operations means  within a DGM network one can simulate, with arbitrary accuracy, any information processing task.

\section{Derivation of Eq.~\eqref{qq_eff} \label{sect:derivation}}
We assume a Hamiltonian given by Eq.~\eqref{qq_ham}, with dissipation occurring according to Eq.~\eqref{eqn_AB}. As such we write the full Hilbert space as $\cl{H}=\bigotimes_\alpha\cl{H}_{q,\alpha}\otimes\cl{H}_{\infty,\alpha}$, where the subscript $(q,\alpha)$ refers to the qubit sector in cavity $\alpha=A,B$, and $(\infty,\alpha)$ the corresponding bosonic sector.

Since $H_{AB}$ conserves the number of photons in the joint system, the dissipative term (Eq.~\eqref{diss_AB}), which accounts for leakage, guarantees the final state under evolution of Eq.~\eqref{eqn_AB} is the joint vacuum state.


For simplicity we set $K_0 = 0$. We expand on this comment at the end of this section, and consider the effect of a non-zero $K_0$.
The second term in equation \eqref{qq_ham} only appears at second order in our approximation, hence must be scaled by $1/\sqrt{T}$. Following the tensor ordering of the Hilbert space, we consider the action of the effective Liouvillian, $\cl{L}_{eff}:=-\cl{P}_0\cl{KSKP}_0$, on a state of the form $\rho_{A}\otimes |0\rangle \langle 0| \otimes \rho_B \otimes |0\rangle \langle 0|$, where $\rho_\alpha$ are qubit states.

The action of the first Hamiltonian super-operator ($\cl{K}$) results in:
\begin{widetext}
\be
\label{KP}
\mathcal{KP}_0(\rho) = -i\sum_{\alpha=A,B}g_\alpha\left(S^-_\alpha\rho_{q,\alpha}\otimes |1\rangle_\alpha\langle 0| - \rho_{q,\alpha}S^+_{\alpha,q}\otimes |0\rangle_\alpha\langle 1|\right),
\ee
\end{widetext}
where identity operations have been ignored for clarity. We define $\hat K := i\cl{KP}_0(\rho)$.
 As in \cite{zanardi-dissipation-2014,dissi_2nd_order:Zanardi2016}, we calculate $\cl{S}$ using the integral form $\cl{S}=-\int_0^{\infty}dt e^{t\mathcal{L}_0}\cl{Q}_0$, where $\cl{Q}_0=1-\cl{P}_0$. Since $\hat K$ has already been projected out of the SSS, we just need to consider the action of $e^{t\cl{L}_0}$, which fortunately can be simplified, noting that $\cl{H}_{AB}:=-i[H_{AB},\cdot]$, and $\cl{L}_{AB}:=\sum_{\alpha=A,B}\cl{L}_0^{(\alpha)}$ commute when acting on $\hat K$. In fact, the action of $\cl{L}_{AB}$ on $\hat K$, or indeed on $\cl{H}_{AB}\hat K$, is to simply pull out a factor of $\frac{-\tau^{-1}}{2}$ (hence the exponential gives $e^{-\frac{t/\tau}{2}}$). Therefore, the task is to calculate $e^{t \cl{H}_{AB}}\hat K$.
 
 Define $\hat K' := \frac{i}{J}\cl{H}_{AB}\hat K$. One can see that $\cl{H}_{AB}\hat K' = -iJ \hat K$ (i.e.~applying $\cl{H}_{AB}$ twice is the identity, up to an overall $-J^2$ factor). Thus, 
\begin{widetext}
\be
e^{t\cl{H}_{AB}}(\hat K) = \hat K-itJ \hat K' -\frac{(tJ)^2}{2!}\hat K+ i\frac{(tJ)^3}{3!} \hat K' + \dots
= \hat K \cos(Jt) -i\hat K'\sin(Jt).
\ee
\end{widetext}
Combining these two results allows us to explicitly perform this integration, and hence calculate $\cl{SKP}_0$. We get:

\begin{widetext}
\be
\begin{split}
\cl{S}\hat K = -\hat K \int_0 ^\infty e^{-\frac{t/\tau}{2}} \cos (Jt) dt + i\hat K' \int_0 ^\infty e^{-\frac{t/\tau}{2}} \sin (Jt) dt \\
= \frac{-2 \tau}{1+4(J\tau)^2}\hat K + \frac{4iJ\tau^2}{1+4(J\tau)^2}\hat K'.
\end{split}
\ee
\end{widetext}

Applying $\cl{K}$ to the above essentially just brings out another factor of $g_A$ or $g_B$, with the appropriate $S^\pm$ operator. Projecting back into the steady state results in
\begin{widetext}
\be
\label{effective_main}
\mathcal{L}_{eff}(\rho) = -i\frac{-4Jg_A g_B \tau^2}{ 1+ 4(\tau J)^2}\left[S^-_A S^+_B + S^+_A S^-_B,\rho\right] 
+ \frac{4\tau}{1 + 4(\tau J)^2}\sum_{\alpha=A,B}g_\alpha ^2 (S_\alpha^- \rho S_\alpha^+ - \frac{1}{2}\{S_\alpha^+S_\alpha^-,\rho\}),
\ee
\end{widetext}
 where the state $\rho$ is a steady state. This is the form as quoted in the main text. Note that if we allow $K_0 \neq 0$, and instead scale this Hamiltonian by $1/T$ (as it has a non-vanishing effective first order term), the above result is the same, with the addition that to the effective Lindbladian, Eq.~\eqref{effective_main}, there is an extra term, $-i[\omega^q_A S^z_A + \omega^q_B S^z_B,\rho]$, where as in the above, $\rho$ is a steady state.

\section{Derivation of Eq.~\eqref{eq:mixed}}\label{sect:derivation_eqMixed}
For a single qubit undergoing dephasing in the $z$-direction, steady states are of the form $\rho_{\text{ss}}(a) := a|0\rangle \langle 0| + (1-a)|1\rangle \langle 1|$ (where $|0\rangle, |1\rangle$ are defined in Sect.~\ref{sect:mixed}). One can easily compute 
\be
\cl{KP}_0(X)= -ig(a-b)[|1\rangle \langle 0|\otimes |0 \rangle \langle 1| - H.c.],
\ee
where $\cl{P}_0(X)=\rho_{\text{ss}}(a) \otimes \rho_{\text{ss}}(b)$, and $\cl{K}$ is defined by Hamiltonian Eq.~\eqref{eq:zHam}. The tensor ordering respects the order of the DGMs. It is clear that this projects to zero at first order (i.e.~$\cl{P}_0\cl{KP}_0=0$), so we consider the second order effective generator.

One can check that 
\be
\cl{L}_{eff}(X)=-\tau g^2 (a-b)[|0\rangle \langle 0| \otimes |1 \rangle \langle 1| - |1\rangle \langle 1| \otimes |0\rangle \langle 0|].
\ee
This allows us to calculate $e^{ T\cl{L}_{eff}}(X)$, for some steady state $X$. In particular, in the main text, we pick $X=|0\rangle \langle 0 | \otimes |1 \rangle \langle 1|$, which results in Eq.~\eqref{eq:mixed}.

\end{document}